\documentclass[aps,prd,preprint,superscriptaddress,nofootinbib]{revtex4-1}
\usepackage{graphicx}
\usepackage{booktabs}
\usepackage{hyperref}
\usepackage{amssymb,amsmath,amsfonts}
\usepackage{color}
\usepackage[utf8]{inputenc}
\usepackage[caption=false]{subfig}

\begin{document}

\title{Finite-temperature phase transitions of third and higher order in gauge theories at large $N$}

\author{Hiromichi Nishimura}
\affiliation{RIKEN/BNL Research Center, Brookhaven National Laboratory, Upton, NY 11973}

\author{Robert D. Pisarski}
\affiliation{Department of Physics, Brookhaven National Laboratory, Upton, NY 11973}

\author{Vladimir V. Skokov}
\affiliation{RIKEN/BNL Research Center, Brookhaven National Laboratory, Upton, NY 11973}
  
\begin{abstract}

We study phase transitions in $SU(\infty)$ gauge theories at nonzero temperature using matrix models.
Our basic assumption is that the effective potential is dominated by double trace terms
for the Polyakov loops.  As a function of the various parameters, related to
terms linear, quadratic, and quartic in the Polyakov loop,
the phase diagram exhibits a universal structure.
In a large region of this parameter space, 
there is a continuous phase transition whose order is larger than second. 
This is a generalization of the phase transition of Gross, Witten, and Wadia (GWW).  
Depending upon the detailed form of the matrix model,
the eigenvalue density and the behavior of the specific
heat near the transition differ drastically.  
We speculate that in the pure gauge theory,
that although the deconfining transition is thermodynamically of first order,
it can be nevertheless conformally symmetric at infinite $N$.

\end{abstract}

\maketitle

\section{Introduction}

The nature of the deconfining phase transition in $SU(N)$ gauge theories is a
question of fundamental importance; numerical simulations
on the lattice indicate a transition of first order for $N \geq 3$ \cite{Lucini:2012gg, *Lucini:2013qja}.
In finite-temperature pure gauge theory, the Polyakov loop is the relevant order parameter. It is therefore reasonable to study the phase transition as a function of an effective theory of thermal Wilson lines, as a type of matrix model.

There are many matrix models which are soluble at large $N$.
The most familiar is when the transition is
driven by the Vandermonde
determinant from the integration measure of a single site integral
\cite{Brezin:1977sv, Gross:1980he, *Wadia:1980cp, Lang:1980ws, *Menotti:1981ry, Jurkiewicz:1982iz,
Green:1983sd, Damgaard:1986mx, Azakov:1986pn,Demeterfi:1990gb,*Jurkiewicz:1990we, Sundborg:1999ue,Aharony:2003sx, *Aharony:2005bq, Dumitru:2003hp, Dumitru:2004gd, AlvarezGaume:2005fv, Schnitzer:2004qt, Hollowood:2009sy, *Hands:2010zp, *Hollowood:2011ep, *Hollowood:2012nr, Ogilvie:2012is, Liu:2015yaa}.
This type of model was originally applied to a lattice gauge theory in two space-time dimensions
\cite{Gross:1980he, *Wadia:1980cp, Lang:1980ws, *Menotti:1981ry, Jurkiewicz:1982iz},
where, for the Wilson action, there is a third-order transition as a function of the coupling constant. The third order transition as a function of temperature was subsequently shown in lattice gauge theory at strong coupling with heavy quarks using the mean-field approximation \cite{Green:1983sd, Damgaard:1986mx}.

Sundborg 
showed that at infinite $N$, this model is relevant to deconfinement on a femtosphere, $S^3 \times R^1$
\cite{Sundborg:1999ue,Aharony:2003sx, *Aharony:2005bq, Dumitru:2003hp, Dumitru:2004gd, AlvarezGaume:2005fv, Schnitzer:2004qt, Hollowood:2009sy, *Hands:2010zp, *Hollowood:2011ep, *Hollowood:2012nr, Ogilvie:2012is}.
As a function of temperature, the deconfining transition appears to be of first order, as both
the energy density and the order parameter are discontinuous at the transition temperature
$T_d$.
Even so, the specific heat
diverges as $T \rightarrow T_d^+$, as is typical of a second order transition
~\cite{Dumitru:2003hp, Dumitru:2004gd, Pisarski:2012bj, *Lin:2013qu, Fukushima:2017csk}.
For this reason the transition at infinite $N$
can be termed ``critical first order'' \cite{Dumitru:2003hp, Dumitru:2004gd}.

On the femtosphere, Aharony {\it et al.} showed that the critical first order is washed out
when higher order perturbative corrections are included, leaving an ordinary
first order transition \cite{Aharony:2003sx, *Aharony:2005bq}.
The question is whether this remains true in the limit of infinite volume.
An effective model of Wilson lines was developed as a model for deconfinement with three colors
\cite{Pisarski:2000eq, *dumitru_degrees_2002,*dumitru_two-point_2002,*scavenius_k_2002, Meisinger:2001cq,  dumitru_dense_2005, *oswald_beta-functions_2006, Pisarski:2006hz, dumitru_eigenvalue_2008, *smith_effective_2013, Dumitru:2010mj, *Dumitru:2012fw, Kashiwa:2012wa, Bicudo:2013yza, *Bicudo:2014cra, Gale:2014dfa, *Hidaka:2015ima}
and extended to include dynamical quarks \cite{Pisarski:2016ixt}.
This model is soluble at large $N$ \cite{Pisarski:2012bj, *Lin:2013qu}.
In this paper we study a general class of matrix models 
and solve them in some special but illustrative cases.

We now give an outline of this paper and summarize the main results. 
Matrix models for the deconfining phase transition are functions of the thermal
Wilson line, $L = \mathcal{P} \exp ( i g \int^{1/T}_0 A_{0} \, d \tau)$, which
we take to lie in the fundamental representation.
The general form of the effective potential, which we discuss in
Sec.~\ref{Effective_potential}, includes an infinity of terms.
The simplest possibility is to start with those involving arbitrary powers of $L$,
but just two traces,
\begin{equation}
  N^2 \, V_{\rm eff} =  \sum^{\infty}_{n=1} a_n  \left| \mathrm{tr} L^n  \right|^2 + \dots \, .
  \label{double_trace}
\end{equation}
We assume that these double-trace terms dominate near the deconfining phase transition $T_d$.  All $a_n$ are therefore positive below $T_d$ to prevent the spontaneous symmetry breaking of $Z(N)$.  

Section \ref{Phase_structure} considers deviations from the double trace terms of Eq.~(\ref{double_trace}).
We assume that the phase transition is driven by traces of loops which wind only once in
imaginary time, 
$\mathrm{tr} \, L$, and not by those which wind more than once,
such as $\mathrm{tr}\, L^2$, {\it etc.}
This is a significant assumption, but is borne out by all known models.  
Then the obvious terms to add next are those linear and quartic in $\mathrm{tr} \, L$:
\begin{equation}
N^2 \,   V_{\rm eff} =  \sum^{\infty}_{n=1} a_n  \left| \mathrm{tr} \, L^n  \right|^2
  + \frac{b_1}{N^2} \left(\left| \mathrm{tr} \, L  \right|^2\right)^2
  - N \, h \left( \mathrm{tr} \, L + \mathrm{tr} \, L^{\dagger}  \right)\; .
\end{equation}
We assume that the coupling for the quartic term, $b_1$, is small near $T_d$, but let the magnitude of
the background field, $h$, be arbitrary.
We ignore all other couplings,
including cubic couplings such as $\mathrm{tr} (L^{-1})^2 (\mathrm{tr} L)^2 + {\rm c.c.}$,
{\it etc.}, and discuss why this might be valid.       

For the original model of Gross, Witten, and Wadia \cite{Gross:1980he, *Wadia:1980cp},
the coefficients $a_n = 1/n$, and there is a phase transition when
the expectation value of the loop, $\frac{1}{N} \mathrm{tr} \, L$, equals $\frac{1}{2}$.
In Sec.~\ref{Phase_structure}, we consider {\it arbitrary} $a_n$, and show that for some critical 
$h_c$, there is {\it always} a phase transition when $\frac{1}{N} \mathrm{tr} \, L $ goes through $\frac{1}{2}$. 
We thus term this point a generalized Gross-Witten-Wadia (GWW) transition.
In Sec.~\ref{Phase_diagram} we show that there is a region in the space of $a_1$, $b_1$ and $h$ where a generalized GWW transition occurs.

About the GWW transition,
the value of the potential at the minimum can be expanded in powers of $\delta h = h - h_c$:
\begin{equation}
F(h) = f_{\rm reg}(h) +
\left\{ 
\begin{array}{lcl}
0 & \mbox{for} & \delta h \leq 0
\\ 
v \delta h^r + \mathcal{O}(\delta h^{r+1})  & \mbox{for} & \delta h > 0  \; ,
\end{array}
\right.
\label{F_intro}
\end{equation}
where $v$ is an irrelevant constant, and
$f_{\rm reg}$ is a smooth function of $h$. 
For the model of Gross, Witten and Wadia, $r=3$, and thus the transition is of third order in $h$.
In Sec.~\ref{Order_GWW} we argue that {\it independent} of the values of the coefficients $a_n$,
$r > 2$
for the generalized GWW transition, and so the transition is of higher order than second.
The point where $h = b_1 = 0$ is special, as lines
of first, second, and higher order meet, Fig.~(\ref{PhaseDiagram_2D}).

We stress that as in the model of Gross, Witten, and Wadia, this unusual phase structure emerges
{\it only} at infinite $N$.
This can be seen from Eq.~(\ref{F_intro}): at infinite $N$ the piecewise
function emerges.  For finite $N$,
however, the corresponding function is regular at $\delta h = 0$, and so
the corresponding transition at $\frac{1}{N} \mathrm{tr} L = \frac{1}{2}$ is
only a smooth crossover.

In Sec.~\ref{Models} we solve a specific class of models.  We take the coefficients that contain a simple form,
\begin{equation}
  a_n \sim \frac{1}{n^s}
\end{equation}
with $s = 1, 2, 3$ and $4$. 
In these instances we solve for the eigenvalue density of the Polyakov loop exactly
and confirm the general analysis in
Sec.~\ref{Phase_structure}. The exponent $r$ in Eq.~(\ref{F_intro}) is computed and equals
\begin{equation}
  r=\frac{5+s}{2} \; 
  \label{eq_relate_r_s}
\end{equation}
when $s=1,2,3$ and $4$. This shows that the GWW transition is of third order when
$s = 1$, as is known, and of higher order when $s = 2,3$ and $4$.
We investigate  how the quartic coupling
changes the behavior of the Polyakov loop near the phase transition.  

In the conclusions, Sec.~\ref{Conclusions}, we discuss the possible implications of
our results.  In particular, we speculate that in infinite volume, that the
deconfining transition can be critical first order.

\section{Effective potential}
\label{Effective_potential}

In this section, we consider the effective potential of the Polyakov loop in
$SU(N)$ Yang-Mills theory at large $N$.
After some standard definitions, we consider the possible forms for the
effective potentials of Polyakov loops, and suggest that 
double trace terms may dominate near the deconfining phase transition.

\subsection{Notation}

In a Yang-Mills theory without dynamical quarks
at nonzero temperature, the global symmetry associated
with the deconfining phase transition is $Z(N)$.
The basic variable is the Wilson loop in the direction of imaginary time, $\tau$,
\begin{equation}
L(\mathbf{x})=
\mathcal{P} \exp \left[ {ig  \int^{1/T}_{0} d\tau A_0 (\tau, \mathbf{x}) }\right] .
\label{Polyakov_Line}
\end{equation}
Under the center symmetry the thermal Wilson line transforms as $L \rightarrow z \, L$
where $z$ is an element of $Z(N)$,
$z = \exp(2 \pi i j/N)$ for an integer $j=1\ldots N$.

While the thermal Wilson line is gauge variant, its eigenvalues are gauge
invariant.  As a unitary matrix, after diagonalization 
\begin{equation}
L(\mathbf{x}) = \mbox{diag} ({\rm e}^{i\theta_1}, \dots, {\rm e}^{i\theta_N} ) \; .
\end{equation}
As an $SU(N)$ matrix, these eigenvalues satisfy
$\sum_{i=1}^N \theta_i = 0$, modulo $2 \pi$.

The phase transition is then characterized by the traces of powers of
the thermal Wilson line.  Without loss of generality we can take all
traces to be in the fundamental representation, so there are $N-1$
independent Polyakov loops, 
\begin{equation}
\rho_n = \frac{1}{N} \; \mathrm{tr}\,  L^n \, .
\label{rho_n}
\end{equation}
The $n^{th}$ Polyakov loop $\rho_n$ wraps around in imaginary time $n$ times.
The $\rho_n$ form a complete set of gauge invariant
order parameters for the spontaneous breaking of $Z(N)$ symmetry
in the deconfined phase \cite{Meisinger:2001cq, Myers:2007vc, *Myers:2009df, *Ogilvie:2007tj}.

At large $N$ we introduce the variable $x$,
\begin{equation}
x = \frac{i}{N} -\frac12 \; ,
\end{equation}
where $\theta_i \rightarrow \theta(x)$ ~\cite{Brezin:1977sv}.  The $n^{th}$ Polyakov
loop is then
\begin{equation}
\rho_n = \frac{1}{N} \sum^N_{i=1} e^{in \theta_i}  \rightarrow  
\int^{\frac{1}{2}}_{-\frac{1}{2}} dx \; {\rm e}^{ i n \theta(x) } \, .
\label{rho_n_x}
\end{equation}
At infinite $N$ each loop is a functional of $\theta(x)$. 
Introducing the eigenvalue density
\begin{equation}
\rho(\theta)= \frac{dx}{d\theta} \, ,
\label{dx/dtheta}
\end{equation}
the loop becomes
\begin{equation}
\rho_n = \int^\pi_{-\pi} d\theta \; \rho(\theta)  \; e^{i n \theta} \; .
\label{rho_n_theta}
\end{equation}
In this way, the Polyakov loops are functionals of $\rho$, rather than of $\theta$.  

The eigenvalue density must be non-negative,
 \begin{equation}
\rho(\theta) \geq 0 \; .
\label{nonnegative_condition}
\end{equation}
This will play an essential role for the GWW phase transition.
It is normalized as
\begin{equation}
\int^{\pi}_{-\pi} d \theta \; \rho(\theta)  = 1 \, . 
\label{normalization_condition}
\end{equation}
Provided that $\rho (\theta)$ is continuously differentiable for $-\pi \leq
\theta \leq \pi$, we can write $\rho$ as a Fourier series in terms of its
moments in $\theta$,
\begin{equation}
\rho(\theta) = \sum^{\infty}_{n=-\infty} \rho_n e^{i n \theta}= \frac{1}{2 \pi} \left( 1 + 2 \sum^{\infty}_{n =1 } \rho_n \cos n \theta \right), \,
\label{rho_Fourier}
\end{equation}
using Eq.~(\ref{normalization_condition}).
We have assumed that the expectation value of every Polyakov loop $\rho_n$ equals its complex
conjugate, $\rho_{-n}$.  They are related under charge conjugation, and so
this assumption is valid at nonzero temperature and zero quark chemical
potential.  (At nonzero chemical potential \cite{Azakov:1986pn} the expectation value of the loop
and its complex conjugate differ
\cite{dumitru_dense_2005,  Hollowood:2009sy, *Hands:2010zp, *Hollowood:2011ep, *Hollowood:2012nr, Ogilvie:2012is, Nishimura:2014rxa, *Nishimura:2014kla, *Nishimura:2015lit}).
In doing so we implicitly perform an overall $Z(N)$ rotation so that the expectation
value of all Polyakov loops are real. 

At infinite $N$, all Polyakov loops vanish in the confined phase,
$\rho_n = 0$ for all $n \geq 1$.  This implies
that the eigenvalue density is constant, $\rho (\theta)= 1/(2\pi)$, demonstrating 
the complete repulsion of eigenvalues.  In the deconfined phase, $\rho_n$ become nonzero.  At infinitely high temperature
all Polyakov loops equal unity; this implies that all eigenvalues become zero, and thus $\rho(\theta)= \delta(\theta)$. 

\subsection{Effective potentials for the Polyakov loops}

The effective potential for the Polyakov loop is constructed formally as \cite{KorthalsAltes:1993ca}
\begin{equation}
\exp \left[-\mathcal{V} T^{d-1} N^2 V_{\rm eff} (\rho_n ) \right]
=
\int DA_\mu e^{-S_{\rm YM} (A_\mu)} \prod^{N-1}_{m=1}
\delta \left(\rho_m - \frac{1}{\mathcal{V} N } \int
d \mathbf{x}   \,\mathrm{tr} L^{m}\right)\, , 
\label{V_eff_def}
\end{equation}
where $S_{\rm YM}$ is the $d$-dimensional Euclidean Yang-Mills action and $\mathcal{V}$ is the
spatial volume.  The right hand side is a path integral over the gauge fields
in the presence of constant background Polyakov loop $\rho_n$. We have scaled
out $\mathcal{V} T^{d-1}$ so that $V_{\rm eff}$ is dimensionless.  By
convention, there is a factor $N^2$, so that $N^2 V_{\rm eff}$ is of order
$N^{0}$ in the confined phase and $N^2$ in the deconfined phase \cite{Thorn:1980iv}. 
The partition function is then given by integrating over the eigenvalues:
\begin{equation}
Z = \int [d \theta] \exp \left[- \bar{N}^2V_{\rm eff} (\rho_n) \right],
\label{Z_YM_V}
\end{equation}
where we define
\begin{equation}
\bar{N}^2=  \mathcal{V} T^{d-1} N^2  \; .
\end{equation}

At the outset, we stress that at infinite $N$
the effective potential is a function of {\it all} loops, for every
$\rho_n$ from $n=1$ to $\infty$.
One might hope to simplify things by considering an effective potential
for just a few loops, such as $\rho_1$, $\rho_2$, {\it etc.}  However,
{\it all} Polyakov loops vanish in the confined
phase, $\rho_n = 0$ for every $n \geq 1$; an effective potential
involving  a finite number of loops cannot force
all loops to vanish.

Alternately one could integrate over all $\rho_n$ for $n \geq 2$, and
construct an effective potential just for $\rho_1$.  This is possible,
but its utility is not apparent to us.  

Using only the global $Z(N)$ symmetry, there are many terms which can appear
in the most general effective potential for the Polyakov loops,
\begin{equation}
  V_{\rm eff}(\rho_n) 
  =  a_{(1,-1)}^{(1,1)} \, \rho_1 \rho_{-1} + 
a_{(2,-2)}^{(1,1)} \rho_2 \rho_{-2} + 
a_{(1,-2)}^{(2,1)} ( \rho_1^2 \rho_{-2} + \rho_{-1}^2 \rho_{2} )
+ a_{(1,-1)}^{(2,2)} \rho_{1}^2 \rho_{-1}^2 
	+ \dots 
\label{general_zn_potential}
\end{equation}
$$
= \sum_{i_1,i_2,...} a_{(i_1,i_2\ldots)}^{(j_1,j_2\ldots)} \; \rho_{i_1}^{j_1} \; \rho_{i_2}^{j_2} \; \ldots
\; ; \; i_1 \, j_1 + i_2 \, j_2 + \ldots = 0 \; ,
$$
where $a^{(2,1)}_{(1,-2)} = a^{(2,1)}_{(-1,2)} $ by charge conjugation symmetry.  
With such a multitude of terms, this effective potential is not of much use.
Thus we discuss some results from perturbative computations, and from
effective models, which suggest that the effective potential relevant at
infinite $N$ may be much simpler.

Consider first the computation of the effective potential in perturbation
theory.  This is a straightforward matter at one loop order
\cite{Gross:1980br,Weiss:1980rj} and has been
carried out to two loop order \cite{Bhattacharya:1990hk,*Bhattacharya:1992qb, KorthalsAltes:1993ca, Dumitru:2013xna,*Guo:2014zra}:
\begin{equation}
a_{(n,-n)}^{(1,1)} \sim   - d_n^{(4)} =  
  - \; \frac{2}{\pi^2} \; \left( 1 - \frac{5 g^2 N}{16 \pi^2} \right) \; \frac{1}{n^4} \; .
  \label{pert_two_loop}
\end{equation}
The notation $d_n^{(4)}$ denotes the deconfined term, computed perturbatively
in four spacetime dimensions.

The simplicity of this result is not obvious.  Na\"ive computation to two loop order
gives a result which is much more 
complicated than that at one loop order \cite{Enqvist:1990}.
After
including a finite renormalization for Polyakov loops
\cite{Belyaev:1991gh,Bhattacharya:1990hk,*Bhattacharya:1992qb, KorthalsAltes:1993ca,
Dumitru:2013xna,*Guo:2014zra},
however, one finds that all terms collapse 
to Eq.~(\ref{pert_two_loop}), just a constant times the result at
one loop order.

What is remarkable about Eq.~(\ref{pert_two_loop})
is that the {\it only} terms which enter involve two traces:
just $i_1 = - i_2$ and $j_1 = j_2 = 1$.  At one
loop order this is automatic, but at two loop order terms with 
three traces can appear, $\sim g^2 N$.  After including the finite renormalization
of the Polyakov loops
\cite{Belyaev:1991gh,Bhattacharya:1990hk,*Bhattacharya:1992qb,KorthalsAltes:1993ca, Dumitru:2013xna,*Guo:2014zra} these vanish. 

A similar computation to
one loop order in $2+1$ dimensions \cite{KorthalsAltes:1996xp} shows
\begin{equation}
a_{(n,-n)}^{(1,1)} \sim -d_n^{(3)} = -\; \frac{1}{2 \pi} \; \frac{1}{n^3} \; .
  \label{pert_pot_3dim}
\end{equation}
For a field in $d$ spacetime dimensions, the corresponding term is
$a_{(n,-n)}^{(1,1)} \sim - 1/n^d$. 
We discuss results to higher loop order in Appendix \ref{AppendixA}.

The sign of the double trace terms in Eqs.~(\ref{pert_two_loop}) and (\ref{pert_pot_3dim})
is negative.  Thus the potential is minimized by maximizing each Polyakov loop,
which is what is expected in the perturbative regime.
To model the transition to a confining phase, it is necessary to add additional
terms.  In Refs. \cite{Dumitru:2010mj,*Dumitru:2012fw}, an effective model
was constructed by adding terms which mimic
deconfined strings in $1+1$ dimensions:
\begin{equation}
  a_{(n,-n)}^{(1,1)} \sim c_n^{(4)} =  \; c_1 \frac{1}{n^2} \; .
  \label{confined_terms}
\end{equation}
We denote this term $c_n^{(4)}$ as that which drives confinement in $3+1$
dimensions.  
This potential can be also derived by adding a mass deformation to gluons \cite{Meisinger:2001cq}. 
We comment that this particular term is driven by detailed
results from numerical simulations on the lattice, especially the presence
of terms $\sim T^2$ in the free energy relative to the usual $\sim T^4$ \cite{Dumitru:2010mj,*Dumitru:2012fw,Meisinger:2001cq}.
The $c_n^{(4)}$  have positive signs, and if
sufficiently large,
drive a transition to a confined phase.

We conclude this section by discussing other evidence for the dominance of double
trace terms in the effective potential for the thermal Wilson line.
This is true in a strong coupling expansion on a lattice, at least to
leading order \cite{DelDebbio:2008ur}.

In super Yang-Mills theories with mass deformations,
it is possible to compute not only the perturbative
contributions to the free energy, but also the dominant non-perturbative
terms \cite{Poppitz:2012sw,*Anber:2014lba,*Anber:2017tug,cherman}.
While it is not obvious \cite{cherman}, it can be shown that in
these models all terms for the effective potential of thermal Wilson
lines only have double traces.

The {\it only} instance of which we are aware
in which terms with three and four traces arise is for gauge theories
on a femtosphere, $S^3 \times R^1$ \cite{Aharony:2003sx,*Aharony:2005bq}.
Explicit computation to three loop order shows there are a variety
of terms, including those with four traces, although they are suppressed
by $\sim g^4 N^2$, where the coupling constant $g^2 N$ is small on a femtosphere.

As discussed in Appendix \ref{AppendixA}, in perturbation theory
we expect  $a_{(1,-1)}^{(1,1)} \sim  a_{(2,-2)}^{(1,1)}  \sim g^0$, 
while  $a_{(1,-2)}^{(2,1)} \sim g^4  N^2$
and   $a_{(1,-1)}^{(2,2)} \sim g^4 N^2$. As we shall see, $\rho_1$ is of order 1 near the GWW phase transition, 
while $\rho_2\sim g^4 N^2$. Therefore the term   
$a_{(1,-2)}^{(2,1)}  \rho_1^2 \rho_{-2} \sim g^8 N^4 $
can be neglected to the order 
$g^4 N^2 $ , while 
$a_{(1,-1)}^{(2,2)} \rho_1^2 \rho_{-1}^2 \sim g^4 N^2 $
is to be kept. 
A similar quartic term involving $\rho_{n>1}$
will be of order $g^8 N^4 $ or higher and thus will not be included in our analysis.

Consequently, in the following we assume that any terms with three and more traces are small, and compute about that limit.  In the conclusions, Sec.~\ref{Conclusions},
we consider the implications if there are only double trace terms
in the limit of infinite volume.

\section{Phase structure}
\label{Phase_structure}

Motivated by the above considerations, we are led to consider
the following effective potential for a gauge theory at infinite $N$:
\begin{equation}
  V_{\rm eff} =  \sum^{\infty}_{n=1} a_n  \left| \rho_n \right|^2
  + b_1  \left( \left| \rho_1 \right|^2 \right)^2 -  h  \left( \rho_1 +\rho^{*}_1 \right)  \; .
\label{V_eff_h}
\end{equation}
We assume that the deconfining transition at a temperature $T_d$ defined at $h=0$  
is driven by the first Polyakov loop.
In the confined phase, that the first Polyakov loop vanishes
does not necessarily imply that all other Polyakov loops vanish.
For this reason, as discussed above it is essential that we include {\it all}
Polyakov loops in the effective potential and require that all $a_n$ are positive below $T_d$.
We further assume that
\begin{equation}
a_n >0 \,\,\,\,\, \mbox{for} \,\,\,\,\, n \geq 2
  \label{keep_sign_an}
\end{equation}
near $T_d$, so the higher corrections for $\rho_n$ with $n \geq 2$ are not necessary. 

We include a quartic term for the first Polyakov loop, with a coupling $b_1$
which we assume is small and constant near $T_d$.  
We also include a background field for the first Polyakov loop, $\sim h$.
This is natural to include in any spin model, and we can analyze the
model for arbitrary values of $h$.  A background field for 
Polyakov loops is generated by the coupling to quarks \cite{Schnitzer:2004qt}.
If the quarks are heavy, with a mass $m$, then the background field
is $\sim \exp(- m/T)$ for the first Polyakov loop, $\sim \exp(-2 m/T)$
for the second, and so on.  Thus very heavy quarks only
generate a background field for the first Polyakov loop.
For $N_f$ flavors of quarks $h \sim N_f/N \exp(- m/T)$, so we need to
take $N_f \sim N$ for $h$ to persist at infinite $N$.

\subsection{Phase diagrams}
\label{Phase_diagram}

In this subsection we derive general conclusions about the phase diagram of Eq.~(\ref{V_eff_h}) with the condition Eq.~(\ref{keep_sign_an}) near $T_d$. 
In Sec.~\ref{Models} we then solve the series of models
where the $a_n$ have specific values, $a_n \sim 1/n^s  $ for $s = 1,2,3$ and $4$.

At infinite $N$ we look for a saddle point of $V_{\rm eff}(\rho_n)$ under the constraints of
Eqs.~(\ref{nonnegative_condition}) and (\ref{normalization_condition}).
By a $Z(N)$ rotation we can assume that the expectation value of all
Polyakov loops are real, so
$\rho^*_n = \rho_{-n} = \rho_n $.
Na\"ively, the saddle point corresponds to the minimum of each free energy for $\rho_n$,
\begin{equation}
  \frac{d}{d \rho_n} \; V_{n} (\rho_n) \; = 0
  \; 
\end{equation}
where $V_n$ is defined as $V_{\rm eff} = \sum^{\infty}_{n=1}V_n (\rho_n)$.  
It is easy to solve this equation, taking all loops beyond the first to vanish,
$\rho_n = 0$ for $n \geq 2$.  The eigenvalue density in Eq.~(\ref{rho_Fourier})
is then a sum of a constant and $\rho_1$,
\begin{equation}
\rho(\theta) = \frac{1}{2 \pi} (1+ 2 \, \rho_1 \, \cos \theta) 
\; , \;  -\pi  \leq \theta \leq \pi\, .
\label{rho_belowGWW}
\end{equation}
This satisfies the normalization condition of Eq.~(\ref{normalization_condition}),
but it is non-negative only if the first Polyakov loop is
less than or equal to one half, $\rho_1 \leq \frac{1}{2}$.
Therefore, this solution is valid only for  $0 \leq \rho_1 \leq \frac{1}{2}$.

This is the simplest way to see that the point where the first Polyakov loop
equals one half and all others vanish, $\rho_1 = \frac{1}{2}$ and $\rho_n = 0$ for
$n \geq 2$, is special.  We call this the Gross-Witten-Wadia (GWW) point, and
the locus of such points is a GWW surface.

When the expectation value of the first Polyakov loop is greater than $\frac{1}{2}$,
expectation values for all higher loops develop.  This is not due to the usual manner
of Landau mean field, through the coupling of $\rho_1$ to the other $\rho_n$ through
terms such as $(\rho_1^*)^2 \rho_2$, {\it etc.}  
Instead, the eigenvalue density becomes no longer continuously differentiable due to the non-negativity constraint, and as a result higher Polyakov loops become nonzero. In the model of Gross, Witten, and Wadia \cite{Gross:1980he,Wadia:1980cp}, and for the models of Sec.~\ref{Models}, this happens by developing a gap in the eigenvalue density.

\begin{figure}[t]\centering
  \begin{minipage}{.47\textwidth}
    \includegraphics[scale=0.5]{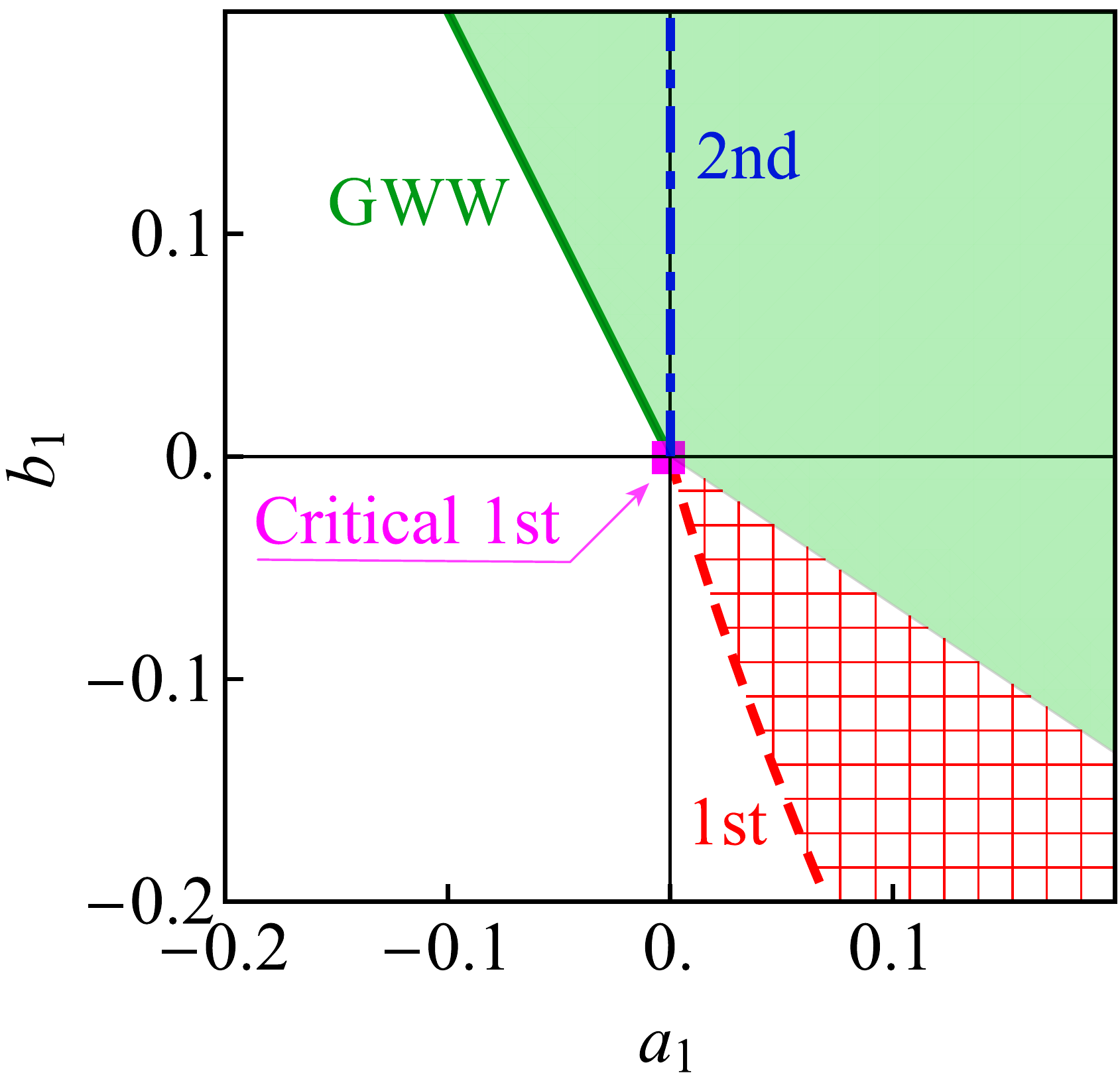}
  \end{minipage}\; \;
  \caption{
    Phase diagrams for the matrix model in Eq.~(\ref{V_eff_h}), this figure and
    Fig.~(\ref{figs_h_a1}).
    This figure shows zero external field, $h=0$, varying the mass
    term, $a_1$, and the quartic coupling, $b_1$, for the first
    Polyakov loop, $\rho_1$.
    The red dashed line is a first order transition; the blue dash-dotted line is a second order transition; the green solid
    line is for the generalized Gross-Witten-Wadia (GWW) transition. The critical first order is located at the
    origin where all three phase transition lines meet.
    For illustration we use
    the model with a Vandermonde determinant for the red dashed line, $s=1$ in Sec.~\ref{Models}.
    The confined phase is the region to the right of the red dashed and blue dash-dotted
    lines, the deconfined to the left of the lines. The green shaded and red hatched regions are the projections of the surfaces of the GWW and first-order phase transitions onto the $h=0$ plane, respectively.
  }
  \label{PhaseDiagram_2D}
\end{figure}

If the first Polyakov loop has an expectation value less than $\frac{1}{2}$,
we can use an effective theory for just that loop, $\rho_1$:
\begin{equation}
V_1 = a_1 \rho^2_1  +b_1  \rho^4_1 -   2 h \rho_1 .
\end{equation}
Consider first zero external field, $h=0$, as illustrated 
in Fig.~(\ref{PhaseDiagram_2D}). If $a_1$ and $b_1$ are positive, the minimum
is clearly for $\rho_1 = 0$.  If $a_1$ is negative and $b_1$ positive,
the minimum is $\rho_1 = \sqrt{-a_1/(2 b_1)}$.  Thus there is a second
order phase transition when $a_1$ vanishes.
This is indicated by the blue dash-dotted line in Fig.~(\ref{PhaseDiagram_2D}).

As $a_1$ decreases for a fixed positive value of $b_1$, the Polyakov loop equals $\frac{1}{2}$ when
$b_1= - 2 a_1$.  At this point, it is no longer possible to include only the first
Polyakov loop in the effective theory.  This is denoted by the green solid GWW line in Fig.~(\ref{PhaseDiagram_2D}).

For negative  $b_1$
we expect a first-order phase transition at some
$a_1 >0$ \cite{Aharony:2003sx,Dumitru:2004gd,AlvarezGaume:2005fv,Fukushima:2017csk}.
The location of the first-order phase transition depends on the explicit form of $a_n$.  
The red dashed line in Fig.~(\ref{PhaseDiagram_2D}) corresponds to the model based on the
Vandermonde determinant ($s=1$) in Sec.~\ref{Models}.

At the origin $a_1= b_1 = h = 0 $, the first, second and higher order phase transition lines meet.  At this point, the
Polyakov loop $\rho_1$ jumps from $0$ to $1/2$, as is typical of a first order phase transition, while
the mass associated with $\rho_1$ becomes zero, as is typical of a second order phase transition. This
point was termed as ``critical first order" in Refs.~\cite{Dumitru:2003hp,Dumitru:2004gd}. 

\begin{figure}[t]\centering
\begin{minipage}{.47\textwidth}
  \subfloat[
  Zero quartic coupling, $b_1=0$.]{
\includegraphics[width=\textwidth]{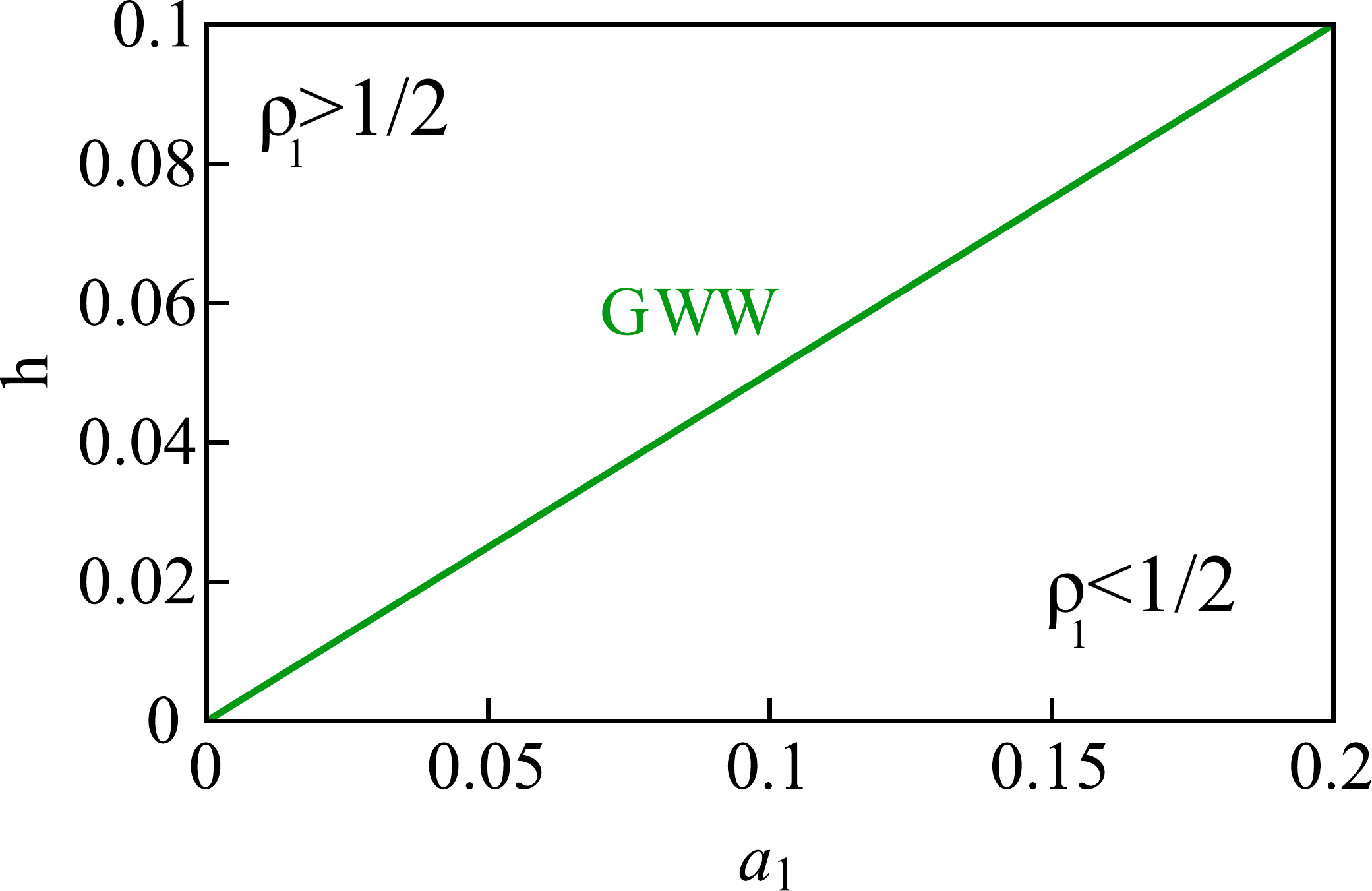}
\label{PhaseDiagram_zero_b}
}\end{minipage}
\begin{minipage}{.47\textwidth}
\subfloat[Negative quartic coupling, $b_1=-0.08$.]{
\includegraphics[width=0.97\textwidth]{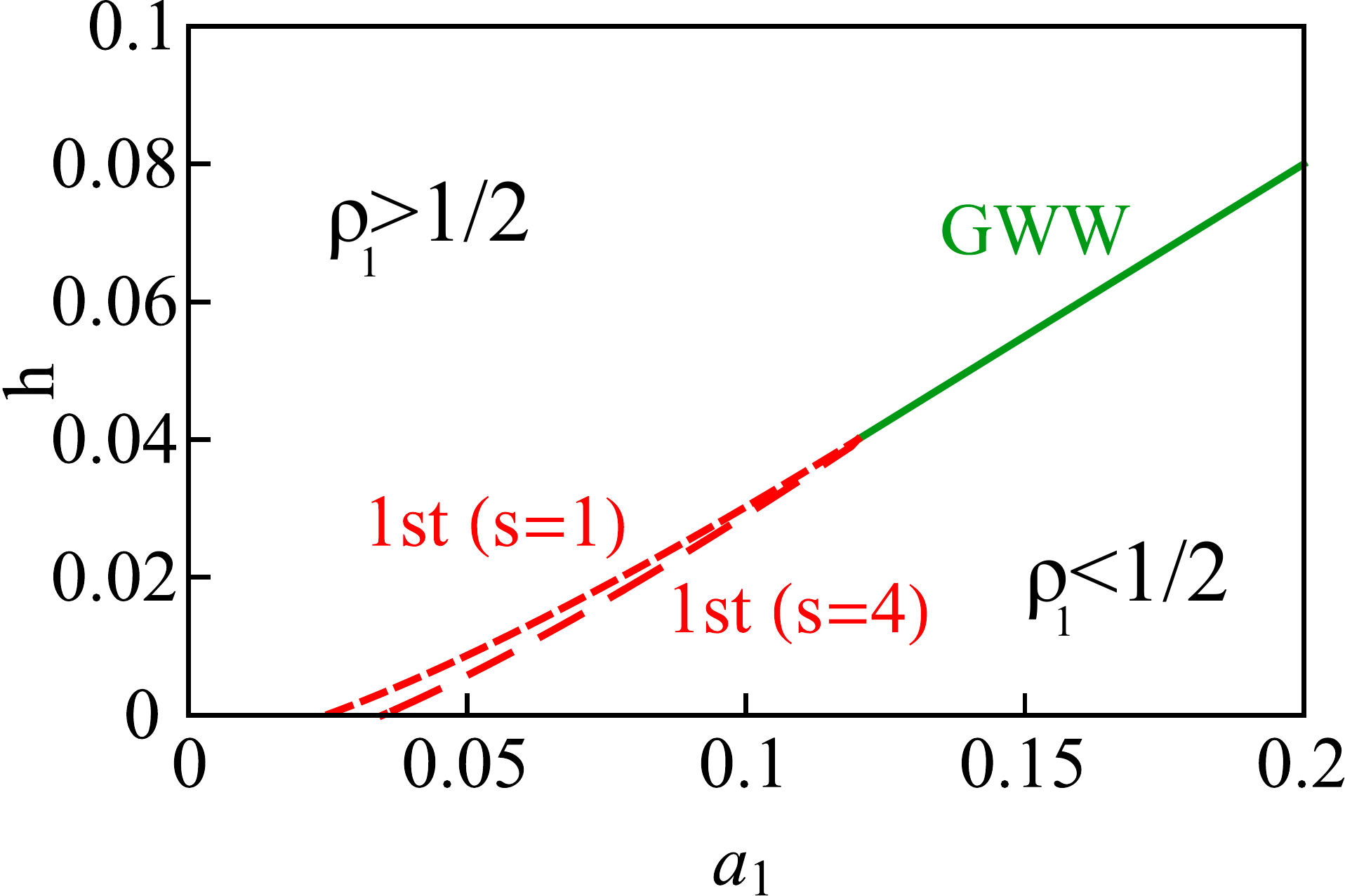}
\label{PhaseDiagram_neg_b}
}\end{minipage}
\caption{
  As in Fig.~(\ref{PhaseDiagram_2D}), versus the background field $h$ and the quadratic
  coupling $a_1$.  We also illustrate the (small) difference between $s=1$ and $s=4$
  in the figure on the right hand side.  
}
\label{figs_h_a1}
\end{figure}

The center symmetry is broken by a nonzero background field $h \neq 0$, which
thus washes out a second order phase transition.  About $\rho_1 = \frac{1}{2}$, we introduce
\begin{align}
  &\delta \rho_1 = \rho_1 - \frac{1}{2} \; , \label{V_1} \\ 
&V_1 = \frac{1}{16} \left(4 a_1 +b_1 - 16 h \right) + \left(a_1 +\frac{b_1}{2} -
2 h \right) \delta \rho_1 +  \left(a_1 +\frac{3 b_1}{2} \right)  \delta
  \rho^2_1 +2 b_1 \delta \rho^3_1 + b_1 \delta \rho^4_1\,
  \notag
\end{align}
where $-1/2 \leq \delta \rho_1 \leq 0$.
This is equivalent to the
Legendre transform $\Gamma(\rho_1)$ of the effective potential below the GWW
point.  As we argue in general in Sec.~\ref{Order_GWW}, and show explicitly in
Eq.~(\ref{Gamma_GWW_model}), the GWW point is a continuous phase
transition, whose order is always higher than second.
Consequently, the coefficients up to and including $\delta
\rho^2$ are continuous about the GWW point.

We analyze Eq.~\eqref{V_1} as follows.  At the GWW point 
$\delta \rho_1=0$, and two conditions need to be satisfied.
First, the
coefficient of $\delta \rho_1$ must vanish, so that $a_1 + b_1/2 - 2 h=0$;
second, that the coefficient of $\delta \rho^2_1$ must be  positive, $a_1 +  3 b_1/2>0$. This forms a surface of GWW points in the space of $a_1, b_1,$ and $h$. The green shaded region in Fig.~(\ref{PhaseDiagram_2D}) indicates the projection of the GWW surface onto the $h=0$ plane.  
The GWW surface is independent of the coefficients $a_n$.
The green solid lines in Fig.~(\ref{figs_h_a1}) are the cross-sections of the GWW surface for $b_1=0$ and $-0.08$.

As $b_1$ is decreased for a fixed, positive value of $a_1$ along the GWW surface, we eventually
hit the boundary where the coefficient of $\delta \rho^2_1$ vanishes, $b_1 = -2 a_1/ 3$.
Beyond this point, $\rho_1=\frac{1}{2}$ is an unstable solution, and there
is a first order transition. This is indicated by a red dashed line
in Figs.~(\ref{PhaseDiagram_2D}) and (\ref{figs_h_a1}).
The red hatched region in Fig.~(\ref{PhaseDiagram_2D}) is the projection of the surface of first order phase transitions onto the $h=0$ plane. 
The location of the first order lines
depends on explicit values of the
$a_n$.  In Fig.~(\ref{PhaseDiagram_neg_b}) we show the lines for $s=1$ and $s=4$: as can
be seen, they are not very different.
The lines for $s=2$ and $3$ lie somewhere between the lines for $s=1$ and $4$.  
 
The Polyakov loop $\rho_1$ becomes larger than $1/2$ above the first order phase transition or GWW point.  In this region, the effective potential is not just a function of the first Polyakov loop, $\rho_1$, but of all $\rho_n$.  To describe the
theory beyond the GWW point, we need to know the explicit values of the coefficients
$a_n$.  We can show, however, that the GWW point is a phase transition point
for arbitrary $a_n$.

\subsection{The order of phase transition at the GWW point}
\label{Order_GWW}

In this subsection we argue that about the GWW point,
there is a continuous phase transition whose order is always higher than second.
The partition function in Eq.~(\ref{Z_YM_V}) can be written as
\begin{equation}
Z = \int [d \theta] e^{- \bar{N}^2 V_{\rm eff} (\rho_n)}
= e^{- \bar{N}^2 F(h) }\,  , 
\label{Z_F}
\end{equation}
where $F$ is the dimensionless free energy in the presence of the external field $h$ per
volume $\mathcal{V}$ and per the color degrees of freedom $N^2$.
$F$ is a generating function for the Polyakov loop, where
the expectation value of $\rho_1$ 
is given by
\begin{equation}
  \frac{d F}{d h}  = - 2 \rho_1(h) \; ;
\end{equation}
the factor of $2$ accounts for the complex conjugate of the first Polyakov loop.
Consequently, the free energy is the integral of the loop with respect to $h$,
\begin{equation}
F(h) = - \, 2 \, \int^{h}_{0} dh' \rho_1(h') \; .
\label{F_GWW_0}
\end{equation}
Since $F$ is the value of the potential at a saddle point of $V_{\rm eff}$ at large $N$,  we have
\begin{equation}
F(h)
= V_{\rm eff} (\rho_n (h) ) \; ,
\label{F_GWW_2}
\end{equation}
where the $n^{th}$ Polyakov loop, $\rho_n(h)$,
satisfies the equation of motion,
\begin{equation}
  \frac{\delta}{\delta \theta(x)} V_{\rm eff} = 0 \; .
\end{equation}
We now have two expressions for the free energy, Eqs.~(\ref{F_GWW_0}) and (\ref{F_GWW_2}). For completeness we give another form of
the free energy in Appendix \ref{AppendixB} when $V_{\rm eff}$ is given as in Eq.~(\ref{V_GWW_Sum}). 

In order to explore the order of phase transition about the GWW point, we only 
need to look at one point in the green shaded region in 
Fig.~(\ref{PhaseDiagram_2D}). We choose the point where the quartic
coupling vanishes and the quadratic couplings are positive, 
\begin{equation}
b_1=0 \;\;\;\;\; \mbox{and} \;\;\;\;\;  a_n > 0 
\end{equation}
for all $n \geq 1$.  
Without loss of generality we choose $a_1=1$.  The effective potential becomes
\begin{equation}
V_{\rm GWW} (\theta ) 
\equiv
 \sum^{\infty}_{n=1} a_n \, \rho^2_n  - 2 \, h \, \rho_1  \, . 
\label{V_GWW_Sum}
\end{equation}
We call it the GWW potential.  This potential naturally appears after 
the Legendre transform of the full potential as shown in Sec.~\ref{Model_setup}.  
The equation of motion is
\begin{equation}
h \sin \theta = \sum^{\infty}_{n=1} n \, a_n \, \rho_n \, \sin (n \theta) \; ,
\label{EoM_GWW_1}
\end{equation}
by using $\delta \rho_n /\delta \theta = - n \sin (n \theta)$ from Eq.~(\ref{rho_n_x}). 
As we discussed in the previous subsection, the equation of motion
(\ref{EoM_GWW_1}) below the GWW point is satisfied if
\begin{eqnarray}
\rho_1 = h
\;\;\;\;\; , \;\;\;\;\;
\rho_{n}=0 \;\; , \;\; n \geq 2 \; .
\end{eqnarray}
By Eq.~(\ref{rho_Fourier}), 
\begin{equation}
\rho = \frac{1}{2\pi } (1 + 2  h \cos \theta)  \,\,\,\,\, \mbox{with} \,\,\,\,\, -\pi  \leq \theta \leq \pi 
\label{rho_belowGWW_h}
\end{equation}
when $0 \leq h \leq \frac{1}{2}$. The GWW point is when
$h=\frac{1}{2}$.  We write the Polyakov loop just below the GWW point
as $\rho_1 = \frac{1}{2} + \delta h$ where $\delta h = h -\frac{1}{2}$. Note that $\delta h$ is
negative below the GWW point. Using Eqs.~(\ref{F_GWW_0}) or (\ref{F_GWW_2}),
the free energy for the GWW potential $V_{\rm GWW}$ is
\begin{equation}
  F_{\rm GWW} = -h^2=  -\frac{1}{4} -\delta h -\delta h^2 \; , \;
  -\frac{1}{2} \leq \delta h \leq 0 \; .
\label{F_belowGWW_h}
\end{equation}

Next consider just above the GWW point, $h = \frac{1}{2}+\delta h$ with $1 \gg \delta h>0$.
Writing the Polyakov loop as $\rho_1 = \frac{1}{2} + \delta \rho_1$,
the equation of motion is
\begin{equation}
0 =\left( \delta \rho_1 -\delta h \right) \sin \theta + \sum^{\infty}_{n=2} n \, a_n \, \rho_n \, \sin (n \theta) \; .
\label{EoM_GWW_2}
\end{equation}
At small $\delta h > 0$ the leading term for the first Polyakov loop is
\begin{equation}
\delta \rho_1  \sim  u \; \delta h^q 
\label{rho1_u}
\end{equation}
where $u$ and $q$ are some constants, with $q \geq 0$.
From Eqs.~(\ref{F_GWW_0}) and (\ref{F_GWW_2}), 
\begin{eqnarray}
F_{\rm GWW}
&\sim& 
-\frac{1}{4} - \delta h - \frac{2u}{1+q} \, \delta h^{1+q} \; ,
\label{F_NearGWW_0}
\\
F_{\rm GWW}
&\sim& 
-\frac{1}{4} -\delta h + u^2 \, \delta h^{2q} -2 \, u \, \delta h^{1+q}  + \sum^{\infty}_{n=2} a_n \, \rho^2_n \; ,
\label{F_NearGWW_2}
\end{eqnarray}
provided that $\delta \rho_1 \sim u \, \delta h^q$.
If $\rho_1 = \frac{1}{2}$ and thus $\delta \rho_1 =0$  above the GWW point,
then the two expressions for the free energy are equal only if
$\sum^{\infty}_{n=2} a_n \rho^2_n $ vanishes. This implies that
all higher Polyakov loops vanish, $\rho_{n} =0$ for $n \geq 2$, which
violates the equation of motion in Eq.~(\ref{EoM_GWW_2})
when $\delta h>0$. Therefore $\rho_1 \neq \frac{1}{2}$ above the GWW point. 
On the other hand, if $u$ is nonzero and $0 \leq q <1$, then we can compare
Eqs.~(\ref{F_NearGWW_0}) and (\ref{F_NearGWW_2}) to obtain
\begin{equation}
\sum^{\infty}_{n=2} a_n \, \rho^2_n \sim - u^2 \, \delta h^{2q} \; .
\end{equation}
This is not consistent, because the $a_n$ are positive and the
$\rho_n$ are real.  Therefore
\begin{equation}
  q \geq 1 \; .
\end{equation}
Comparing Eqs.~(\ref{F_belowGWW_h}) and
(\ref{F_NearGWW_0}) just below and above the GWW, we see that
only the second or higher derivatives of the free energy
are discontinuous.  Hence the phase transition is of second or
higher order.

We now exclude the possibility of a second order transition. If $q = 1$ and $u\neq 1$, or $q>1$,
then the second derivative
of the free energy is discontinuous at the GWW point.
This implies that the mass for the first Polyakov loop is
discontinuous at the GWW point.  The mass for the $n^{th}$ Polyakov loop below and at the GWW point is 
\begin{equation}
\frac{d^2 \; }{ d \rho_n d\rho_m } \; V_{\rm GWW} = 2 \, a_n \, \delta_{m,n} \; .
\end{equation}
This is a diagonal matrix whose elements are
nonzero.  Therefore, if the phase
transition is of second order, any mass eigenvalue
is {\it non}zero at the GWW point.  This is not
expected for a second order transition, where the critical fields
are massless.

The remaining possibility is that $q= u = 1$. Then $\delta \rho_1 \sim \delta h$,
and to leading order, the first term in Eq.~(\ref{EoM_GWW_2}) vanishes.
By comparing Eqs.~(\ref{F_belowGWW_h}) and (\ref{F_NearGWW_0}), the free energy
is continuous up to $\delta h^2$ near the GWW point.  
If we consider the term at next to leading order for $\delta \rho_1$ above the GWW point and use the same argument as before, 
\begin{equation}
  F_{\rm GWW} (h) = - \frac{1}{4} - \delta h - \delta h^2+ 
  \left\{ 
\begin{array}{lcl}
0 & \mbox{for} & \delta h \leq 0
\\ 
v \delta h^r + \mathcal{O}(\delta h^{r+1})  & \mbox{for} & \delta h > 0  \; ,
\end{array}
\right.
  \; ,
\label{F_nearGWW}
\end{equation}
where $v$ is a nonzero constant.  Hence
\begin{equation}
  r > 2 \; .
\end{equation}
This implies that the order of the transition is higher than second.
This is valid for the explicit solutions in Sec.~\ref{Models}.

\section{Models}
\label{Models}

We now solve certain models with special values for 
the coefficients $a_n$ in Eq.~(\ref{V_eff_h}).
We confirm the general phase structure discussed in the previous section,
and compute how the behavior of the Polyakov loops and thermodynamic quantities change.

\subsection{Special cases}
\label{Model_setup}

We consider a simple class of models which are exactly soluble at large $N$:
\begin{eqnarray}
V_{\rm eff} 
   =  
  c_1   \sum^\infty_{n=1} \frac{1}{n^s}  \left| \rho_n\right|^2
  - d_1 \left| \rho_1 \right|^2
  + b_1 \left(\left| \rho_1 \right|^2\right)^2
  -  h \left( \rho_1 + {\rm c.c.} \right) \; ,
\label{V_eff_model}
\end{eqnarray}
with $s=1,2,3$ and $4$.
$\rho_n$ is the $n^{th}$ Polyakov loop, Eq.~(\ref{rho_n}).
This is to take the positive and negative parts of the coefficients, $a_n = c_n - d_n$, in Eq.~(\ref{V_eff_h}) as
\begin{equation}
  c_n = c_1 \frac{1}{n^s} 
  \;\;\;\;\; \mbox{and}   \;\;\;\;\;
  d_n = d_1 \delta_{1 n} \,,
\end{equation}
where $c_1$ and $d_1$ are dimensionless positive functions of $T$. The coefficients $c$ and $d$ denote the ``confined" and ``deconfined", respectively, because they are repulsive and attractive potentials for the eigenvalues as mentioned in Sec.~\ref{Effective_potential}.  

Using the identity
\begin{equation}
  \sum^{\infty}_{n=1}  \frac{\cos (n \phi )}{n^s}   = \frac{1}{2}
  \left( \mathrm{Li}_s (e^{i \phi})+\mathrm{Li}_s (e^{-i \phi} ) \right) \; ,
\label{identity_s=1}
\end{equation}
we see that the effective potential involves the polylogarithms of order $s$.
When $s$ is an even integer, the polylogarithm simplifies further to a Bernoulli
polynomial $B_s$,
\begin{equation}
 \sum^{\infty}_{n=1} \frac{\cos(n  \phi )}{n^{s}} = 
 \frac{(-1)^{\frac{s}{2}+1} (2 \pi)^{s}}{2 \, s!} \;
 B_{s}\left(\frac{\left| \phi \right|}{2 \pi}\right) \; 
 \label{s_even_Fourier}
\end{equation}
where $-2 \pi \leq \phi \leq 2 \pi$.  
For $s=1$, it is related to the Fourier transform of the Vandermonde determinant:
\begin{equation}
\sum^{\infty}_{n=1}\frac{\cos n \phi}{n}  = - \frac{1}{2} \ln \left( 4 \sin^2 \left(\frac{\phi}{2} \right)\right) \; .
\label{s1_Fourier}
\end{equation}
For odd $s$ larger than one, the polylogarithm functions do not simplify further.

To compute thermodynamic quantities at nonzero $b_1$ we perform a Legendre transform \cite{Dumitru:2004gd, AlvarezGaume:2005fv}.
Using the effective potential of Eq.~(\ref{V_eff_model}), 
in the partition function in Eq.~(\ref{Z_YM_V})
we introduce the 
constraint $\delta (\lambda - \rho_1)$,
\begin{eqnarray}
Z 
&=&
\int [d\theta] \int^{\infty}_{-\infty} \frac{d\lambda  d \bar{ \omega } }{2\pi}  \exp \left[ i
\bar{\omega} \left\{ \lambda - \rho_1 \right\}  \right]
\exp\left[ - \mathcal{V} T^{d-1} N^2  V_{\rm eff}  \right]
\\
&=&
\int^{\infty}_{-\infty} \frac{d\lambda  d  \omega  }{2\pi}  
 \exp\left[ -\bar{N}^2 \left( -
 d_1 \lambda^2 +b_1 \lambda^4-  2 h \lambda  + 2 c_1 \omega \lambda \right)
 \right]
Z_{\rm GWW}\, , 
\label{Z}
\end{eqnarray}
where $\bar{N}^2 =  \mathcal{V}T^{d-1} N^2 $ as before, and
$\omega = - i\bar{\omega}/(2 c_1 \bar{N}^2)$ after a Wick rotation.
The delta function constraints the configurations in the path integral to be such that the first Polyakov loop is real.  This is valid because we are only interested in the saddle point where $\rho_n = \rho^*_n$.
We define
\begin{equation}
Z_{\rm GWW} =  \int [d\theta] \exp \left[ -c_1 \bar{N}^2 V_{\rm GWW} \right] = 
 \exp {\left[-c_1 \bar{N}^2 F_{\rm GWW} (\omega) \right]}
\label{Z_GWW_V}
\end{equation}
where
\begin{equation}
V_{\rm GWW} = \sum^\infty_{n=1}  \frac{
\rho^2_n }{n^s}  -2 \, \omega \, \rho_1 \, .
\label{V_GWW_sum_model}
\end{equation}
In Sec.~\ref{Order_GWW} we showed that the second derivative of the free energy, $-d^2 F_{\rm GWW}/d\omega^2$, is positive, and verify this later in
Eq.~(\ref{F_GWW}).  Therefore we can perform a Legendre transform of the free energy $F_{\rm GWW}$ and write the partition function as
\begin{equation}
Z_{\rm GWW} =  \int^{1}_{0} d \rho_1
\exp \left[-c_1 \bar{N}^2 \left\{ \Gamma_{\rm GWW} (\rho_1) - 2 \omega 
\rho_1 \right\} \right]
= 
 \exp {\left[-c_1 \bar{N}^2 F_{\rm GWW} (\omega)\right]} 
\label{Z_GWW_F}
\end{equation}
where 
\begin{equation}
 \left.  \Gamma_{\rm GWW}(\rho_1) =
 \left(
F_{\rm GWW} (\omega) + 2 \omega \rho_1
\right) \right|_{\omega=\omega(\rho_1)} \, .
\label{Gamma_GWW}
\end{equation}
Using Eq.~(\ref{Z_GWW_F}) into Eq.~(\ref{Z}), the total partition function becomes
\begin{equation}
Z = \int^1_0 d \rho_1 \exp \left(- \bar{N}^2 \Gamma(\rho_1) \right)\, ,
\end{equation}
where
\begin{equation}
\Gamma (\rho_1) = - d_1 \rho^2_1+ b_1 \rho^4_1 -2 h \rho_1+ c_1 \Gamma_{\rm GWW} (\rho_1).
\label{Gamma}
\end{equation}
Once we obtain $\Gamma_{\rm GWW}(\rho_1)$ and thus $\Gamma(\rho_1)$,
all thermodynamic quantities can be computed
for given values of $b_1$, $c_1$, $d_1$, and $h$.

\subsection{The GWW potential}
\label{Solution_GWW}

In this subsection we solve for the eigenvalue density $\rho(\theta)$ for the GWW
potential $V_{\rm GWW}$ in Eq.~(\ref{V_GWW_sum_model}). 
The detail derivation is given in the next subsection.
 The potential is
equivalent to Eq.~(\ref{V_GWW_Sum}) by identifying $a_n=1/n^s$ and
taking the background field $h = \omega$.
We solve the model for $s=1,2,3,$ and $4$.

Using Eq.~(\ref{rho_n_x}), the potential can be written in the form 
\begin{equation}
  V_{\rm GWW}=  \int^{\frac{1}{2}}_{-\frac{1}{2}} dx
  \int^{\frac{1}{2}}_{-\frac{1}{2}} dx'  \; \sum^{\infty}_{n=1}
  \frac{1}{n^s} \; \cos\left(n (\theta(x) -  \theta(x'))\right) 
  - 2 \, \omega \int^{\frac{1}{2}}_{-\frac{1}{2}} dx \cos (\theta(x))  \; .
\label{V_GWW}
\end{equation}
The corresponding equation of motion is
\begin{eqnarray}
\omega \sin \theta(x)
&=& 
\int^{\frac{1}{2}}_{-\frac{1}{2}} dx' \sum^{\infty}_{n=1} \frac{1}{n^{s-1}} \, \sin (n (\theta(x)  -  \theta(x')))
\label{EoM_GWW_Integral_1}
\\
&=& 
\int^{\pi}_{-\pi} d\theta' \rho(\theta')  \sum^{\infty}_{n=1}  \frac{1}{n^{s-1}} \, \sin(n (\theta (x) - \theta')) \, , 
\label{EoM_GWW_Integral_2}
\end{eqnarray}
where it is convenient to introduce the eigenvalue density $\rho(\theta)$.  
This is equivalent to Eq.~(\ref{EoM_GWW_1}). It is necessary
to solve the equation of motion under the two constraints of
Eqs.~(\ref{nonnegative_condition}) and
(\ref{normalization_condition}).

\begin{figure}[t]\centering
\begin{minipage}{.47\textwidth}
\subfloat[Below ($\omega = 0, \frac{1}{4}$) and at ($\omega = \frac{1}{2}$) the GWW point.]{
\includegraphics[width=\textwidth]{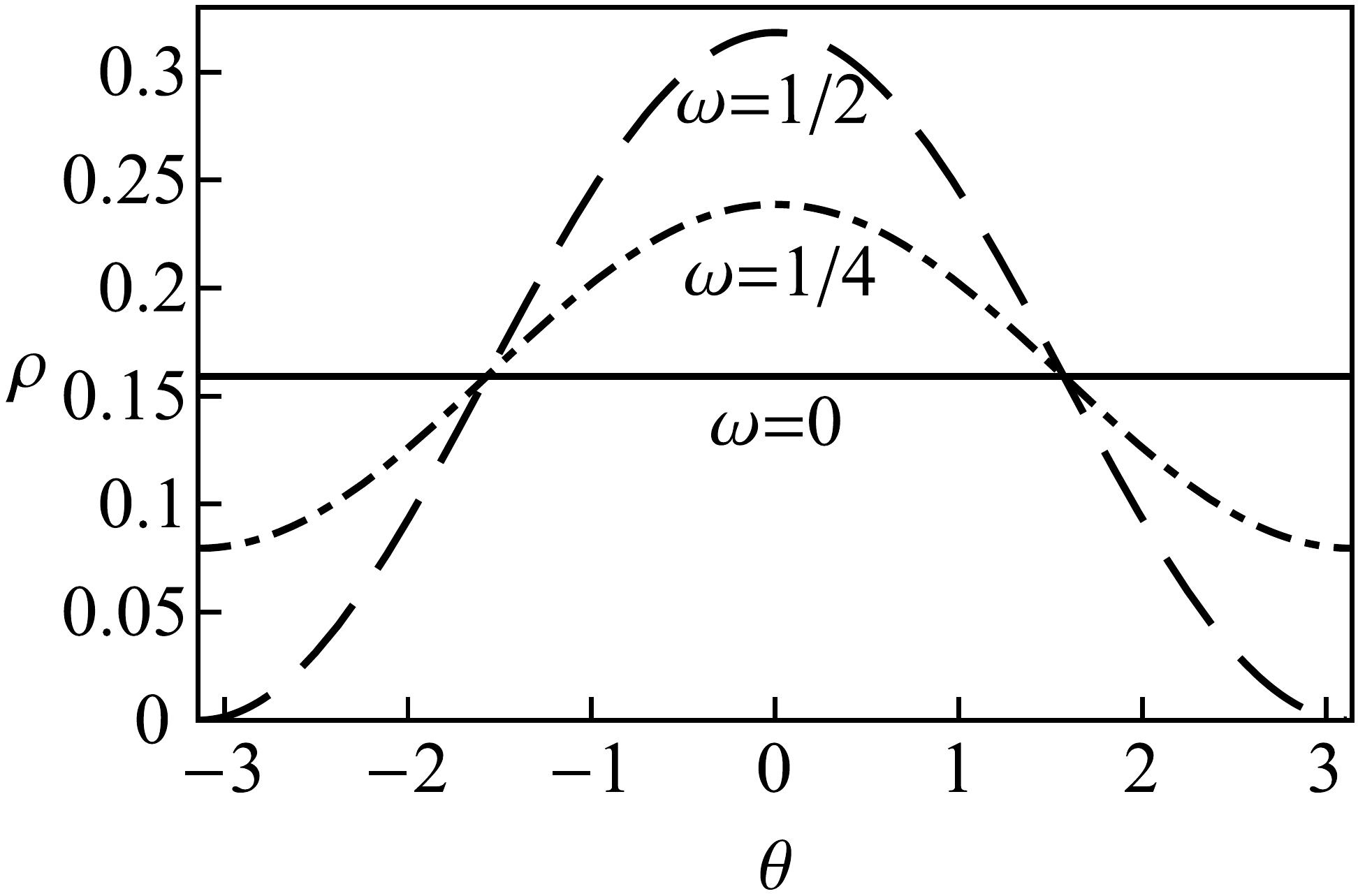}
\label{Distribution_BelowGWW}
}\end{minipage}\; \;
\begin{minipage}{.47\textwidth}
	\subfloat[One point above the GWW point, at $\omega=1$. \vspace{1cm}]{
\includegraphics[width=0.97\textwidth]{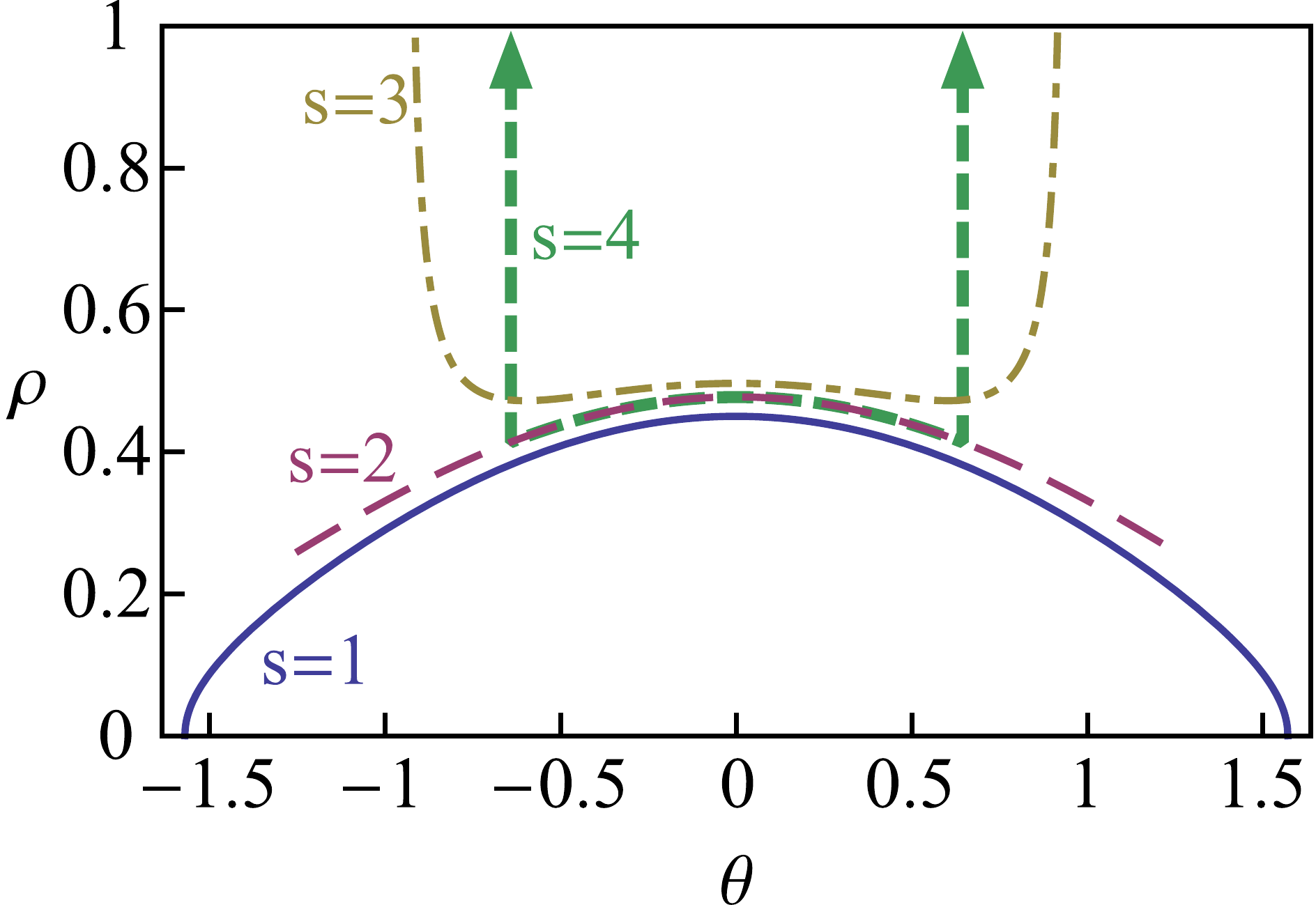}
\label{Distribution_AboveGWW}
}\end{minipage}
\caption{The eigenvalue density as a function of $\omega$. Below and at the GWW
point the density is independent of the model and driven by
the first Polyakov loop, $\rho_1 = \omega$.  Above the GWW point, the density depends on
the coefficients of the double trace terms.}
\label{Distribution}
\end{figure}

For the potential of Eq.~(\ref{V_GWW}), the GWW point corresponds to  $\omega =
\frac{1}{2}$.  Consider first the case below and at the GWW point, where
$\omega \leq \frac{1}{2}$.  It is trivial to solve for the eigenvalue density, 
\begin{equation}
\rho(\theta)=\frac{1}{2\pi} (1 + 2 \omega \cos\theta) \; ;
\label{rho_belowGWW_omega}
\end{equation}
thus $\rho_1=\omega$, and $\rho_{n} =0$ for
$n \geq 2$ in Eq.~(\ref{rho_belowGWW}).
We plot $\rho(\theta)$ for $\omega = 0, \frac{1}{4}$, and $\frac{1}{2}$ in Fig.~(\ref{Distribution_BelowGWW}).
Notice that when $\omega \leq \frac{1}{2}$ the eigenvalue always extends from
$-\pi$ to $+\pi$.  One can also see from Eq.~(\ref{rho_belowGWW_omega}) that this solution
is not consistent for $\omega > \frac{1}{2}$, as then the eigenvalue density
is negative for some range of $\theta$ about $\pm \pi$.

Below the GWW point, $\rho_1= \omega $, so the Legendre transform of $F_{\rm GWW} (\omega) = - \omega^2$  is
\begin{equation}
\Gamma_{\rm GWW}=
\rho_1^2 = \frac{1}{4} + \delta \rho_1 + \delta \rho_1^2 \; ,
\end{equation}
from Eq.~(\ref{Gamma_GWW}), 
where $\delta \rho_1 = \rho_1 - \frac{1}{2} < 0 $.  We write it in this manner
because it will be useful in comparing to the behavior above the GWW point later.  

The properties below and at the GWW point
are independent of the model.  This is not true above the GWW point, and we need to solve the equation of motion for each value of $s$.
In the remaining part of this subsection, we summarize and explain the solutions.

The solutions above the GWW point are given in Eqs.~(\ref{rho_p1}), (\ref{rho_s2}), (\ref{rho_p3}), and (\ref{rho_s4}) for $s=1,2,3,$ and $4$, respectively. For all cases, the eigenvalue density develops a gap at the endpoints, so $\rho(\theta) = 0 $ for $ \theta_0 <  \left| \theta \right|$ where $\theta_0$ is a function of $\omega$ with  $0 \leq \theta_0$. The boundaries
$\pm \theta_0$ of the eigenvalue density are given in Eqs.~(\ref{boundary_p1}),
(\ref{boundary_p2}), (\ref{boundary_p3}), and (\ref{boundary_p4}) for $s=1,2,3$, and $4$, respectively.  At the GWW point, $\theta_0 = \pi$, and the eigenvalue density in all cases becomes the one in Eq.~(\ref{rho_belowGWW_omega}) with $\omega =1/2$. The eigenvalue density is therefore continuous at the GWW transition.

The eigenvalue densities above the GWW point are illustrated in Fig.~(\ref{Distribution_AboveGWW}) for $\omega=1$.  When $\omega > \frac{1}{2}$, a gap opens up in the eigenvalue density for all $s$, so that
it no longer runs from $-\pi$ to $\pi$, but instead from $- \theta_0$ to $\theta_0$.  
The value of $\theta_0$ differs for each value of $s$.
When $s=1$ and $\omega = 1$, the eigenvalue
density runs from $- \frac{\pi}{2}$ to $+ \frac{\pi}{2}$, and vanishes at the ends.
When $s=2$, the eigenvalue density 
jumps discontinuously from a zero to a nonzero value at the ends.
When $s=3$ and $4$, the eigenvalue density actually diverges at the endpoints.

When $s=1$ and $3$ the density has the square root singularities at the endpoints: the density $\rho(\theta)$ for $s=3$ and the derivative $d \rho / d\theta$ for $s=1$ diverge as $\sim 1/\sqrt{\theta_0 - \left| \theta_0 \right|}$ when $\theta \rightarrow \pm \theta_0$.  
This is not surprising for the equation of motion of
Eq.~(\ref{EoM_s1}), which is only well defined by a principal value
prescription.  When $s=2$ and $4$ the density is a simple function of cosine except at the endpoints for $s=4$.  

Physically, as $s$ increases, the change in behavior for the eigenvalue density occurs, because eigenvalue repulsion weakens as $s$ increases.  In particular the value of $\theta_0$ for a fixed value of  $\omega$ becomes smaller as $s$ increases.
When $s=4$ the repulsion is so weak that the eigenvalues pile up at the endpoints $\pm \theta_0$, and the density becomes delta function at the endpoints, as the arrows indicate in Fig.~(\ref{Distribution_AboveGWW}). 
In other words, the eigenvalue $\theta(x)$ for $s=4$ is no longer an injective function of $x$. This implies that there is a critical value of $s=s^{*}$ with $3 < s^* \leq 4$, above which the eigenvalue repulsion is not strong enough to keep all eigenvalues separate above the GWW point.  
This makes the analysis of the case $s \geq 4$ difficult, and 
we do not know if the solution exists above the GWW point in the case $s \geq 5$.  For $s=1,2$ and $3$, all eigenvalues collapse to zero and the density becomes $\rho(\theta) = \delta (\theta)$ as $\omega \rightarrow \infty$.  When $s=4$, all eigenvalues collapse to zero at a finite value of $\omega$ as discussed below Eq.~(\ref{boundary_p4}), again as a consequence of the weak eigenvalue repulsion.

\subsection{Derivation of the solution for the GWW potential}
\label{Derivation_GWW}

In order to solve the equation of motion (\ref{EoM_GWW_Integral_2}) with $s=1,2,3,4$ for the eigenvalue density, we use a trick from Refs.~\cite{Jurkiewicz:1982iz, Pisarski:2012bj}. Let us explain how to solve it when $s$ is a positive integer in a na\"ive way.  
What makes the equation of motion difficult to solve is the combination of
the integral over $\theta'$ and the sum over $1/n^{s-1}$.  However, if we differentiate
with respect to $x$, $\frac{d}{dx} = \frac{d\theta}{dx} \frac{d}{d \theta}$, the sum then becomes $1/n^{s-2}$.  Doing this $s-1$
times, we end up with an integral equation for $\rho(\theta')$, which is soluble.
Each time we differentiate with respect
to $\theta$ we change $\sin(n(\theta-\theta'))$ to $\cos(n(\theta-\theta'))$.  If
we take $s-1$ derivatives, we then end up with a different function depending
upon whether $s$ is even or odd. 

This approach breaks down if $\frac{d\theta}{dx} =0$ for some domains of $x$, i.e.~if eigenvalues pile up.  As mentioned in the previous subsection, the pileup occurs at the endpoints in the case of $s=4$, so we need to treat this case with care.
We first solve for odd $s$, and then even.

\subsubsection{Odd $s=1$ and $3$}

In order to solve the equation of motion (\ref{EoM_GWW_Integral_2}) for $s=3$, we differentiate
it with respect to
$\theta$ twice. Using Eq.~(\ref{s1_Fourier}), we can write the first derivative as
\begin{equation}
\omega \cos \theta = -\frac{1}{2} \int^{\pi}_{-\pi} d \theta' \rho(\theta') \ln \left( 4 \sin^2 \left( \frac{\theta - \theta'}{2}\right)\right) .
\label{EoM_s3_1stDerivative}
\end{equation}
The second derivative is
\begin{equation}
2 \, \omega  \sin \theta = \int^{\pi}_{-\pi} d \theta' \rho(\theta') \cot \left(\frac{\theta - \theta'}{2}\right).
\label{EoM_s1}
\end{equation}
This is the equation of motion for $s=1$, 
and is the circular Hilbert transform of $\rho(\theta')$
with the kernel $\cot(\frac{\theta-\theta'}{2})$
\cite{Henrici:1986:ACC:5664, Muskhelishvili}.  The integral is singular
when $\theta=\theta'$, and so implicitly it is defined by using a principal value
prescription.  The eigenvalue density is the inverse of the transform,
\begin{equation}
\rho (\theta)= C_1 \cos \frac{\theta}{2} \left( \sin^2 \frac{\theta_0}{2} - \sin^2 \frac{\theta}{2} \right)^{\frac{1}{2}} + C_2 \cos^3 \frac{\theta}{2} \left( \sin^2 \frac{\theta_0}{2} - \sin^2 \frac{\theta}{2} \right)^{-\frac{1}{2}} \, , 
\label{rho_odd}
\end{equation}
where the constants satisfy
\begin{equation}
  C_1 + C_2 = \frac{2\omega}{\pi} \; .
  \label{relation_Cs}
\end{equation}
At the GWW point $\theta_0=\pi$, and
the eigenvalue density is that of
Eq.~(\ref{rho_belowGWW_omega}) with $\omega = \frac{1}{2}$. Above the GWW point,
a gap opens up, with the density nonzero only between $-\theta_0$ and $\theta_0$.

For $s=1$, the solution only involves $C_1$ in Eq.~(\ref{rho_odd}), with $C_2 = 0$
\cite{Gross:1980he,*Wadia:1980cp,Jurkiewicz:1982iz}.
The eigenvalue density
above the GWW point, $\omega>\frac{1}{2}$, is
\cite{Gross:1980he,*Wadia:1980cp,Jurkiewicz:1982iz}
\begin{equation}
\rho(\theta)= \frac{2 \omega }{\pi} \cos \frac{\theta}{2} \left( \sin^2
\frac{\theta_0}{2} - \sin^2 \frac{\theta}{2} \right)^{\frac{1}{2}} \, .
\label{rho_p1}
\end{equation}
The endpoint $\theta_0$ is fixed by the normalization condition of Eq.~(\ref{normalization_condition}):
\begin{equation}
\omega = \frac{1}{2 \sin^2 \frac{\theta_0}{2}} \;\;\;\;\; \mbox{or} \;\;\;\;\;
\theta_0 = 2 \sin^{-1}\frac{1}{ \sqrt{2 \omega} } \; .
\label{boundary_p1}
\end{equation}
Using Eqs.~(\ref{rho_n_theta}) and (\ref{rho_p1}), the first Polyakov loop equals
\begin{equation}
\rho_1 (\omega) = 1 - \frac{1}{4 \omega} = \frac{1}{2} + \delta \omega - 2 \delta
\omega^2 + \mathcal{O} (\delta \omega^3) \, , 
\label{rho1_s1}
\end{equation}
where $\delta \omega = \omega - \frac{1}{2}>0$. 

For $s=3$ we need both terms in Eq.~(\ref{rho_odd}) to solve the equation of motion.
Equation (\ref{relation_Cs}) and the normalization condition (\ref{normalization_condition}) give
\begin{equation}
  \rho(\theta) = \frac{1}{2 \pi } \; \cos \left(\theta/2\right) \;
  \frac{1+2 \, \omega \, \left( \sin^2(\theta_0/2) - 2 \sin^2 (\theta/2) \right)}
  {\sqrt{\sin^2(\theta_0/2) - \sin^2(\theta/2) }} \; .
\label{rho_p3}
\end{equation}
We need to determine the position of the endpoint, $\theta_0$, as a function of $\omega$.
This follows from the first derivative of the equation of motion with respect to $\theta$ (\ref{EoM_s3_1stDerivative}). Because the
equation of motion has to satisfy with all values of $\theta$, we can expand to
leading order about $\theta =0$ to find
\begin{equation}
\omega = -\frac{\ln ( \sin (\theta_0/2))}{1- \sin^2 (\theta_0/2)}
\; .
\label{boundary_p3}
\end{equation}
All the higher-order terms turn out to be independent of $\theta_0$.
By introducing the Lambert function or the product logarithm, $W(z)$, defined by the principal solution for 
\begin{equation}
z =   W(z) e^{W(z)} \; ,
\end{equation}
we can invert Eq.~(\ref{boundary_p3}) to obtain $\theta_0$ as a function of $\omega$,
\begin{equation}
\theta_0(\omega) = 2 \sin^{-1}  \sqrt{\frac{ W(-2 \omega e^{-2\omega})}{ -2 \omega } } \, .
\label{boundary_p3_lambert}
\end{equation}

Using Eqs.~(\ref{rho_n_theta}) and (\ref{rho_p3}), the first Polyakov loop equals
\begin{equation}
  \rho_1(\omega) =1- \sin^2 \frac{\theta_0}{2} + \omega \sin^4  \frac{\theta_0}{2} \; .
  \label{rho1_s3}
\end{equation}
Expanding about the GWW point,
\begin{equation}
  \rho_1(\omega) =
  \frac{1}{2} + \delta \omega -\frac{16}{3} \delta \omega^3  + \mathcal{O}(\delta \omega^4) ,
  \label{rho1_s3_expand}
\end{equation}
where $\delta\omega = \omega - \frac{1}{2} > 0$.

\subsubsection{Even $s=2$ and $4$}
\label{solve_even_s}

As with odd $s$, for even $s$ the GWW phase transition
is characterized by a gap in the eigenvalue density.

\begin{figure}[t]\centering
\begin{minipage}{.47\textwidth}
\subfloat[$s=2$ and $N=55$]{
\includegraphics[width=\textwidth]{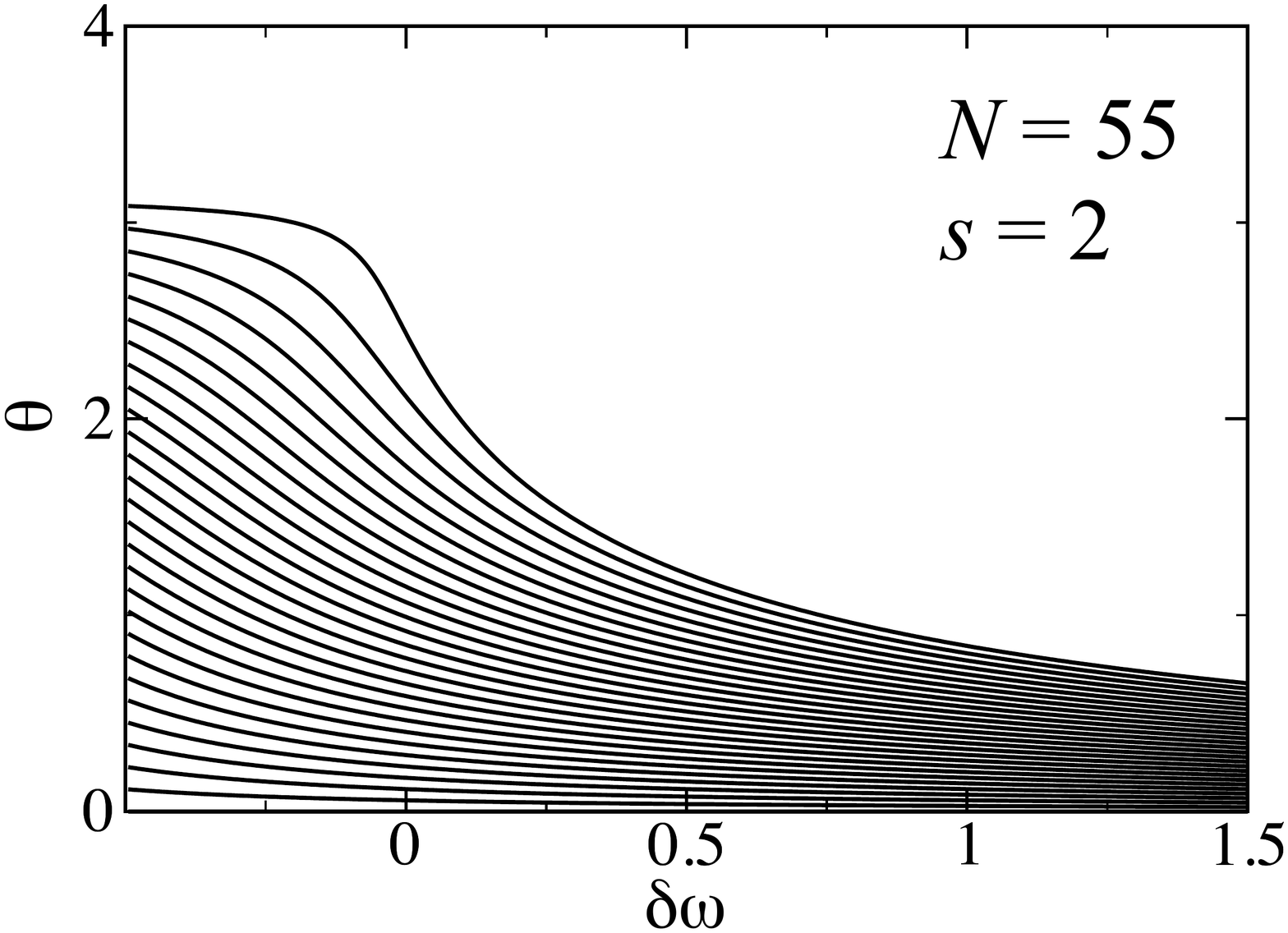}
\label{eigenvalue_finiteN_2}
}\end{minipage}\; \;
\begin{minipage}{.47\textwidth}
	\subfloat[$s=4$ and $N=55$ ]{
\includegraphics[width=\textwidth]{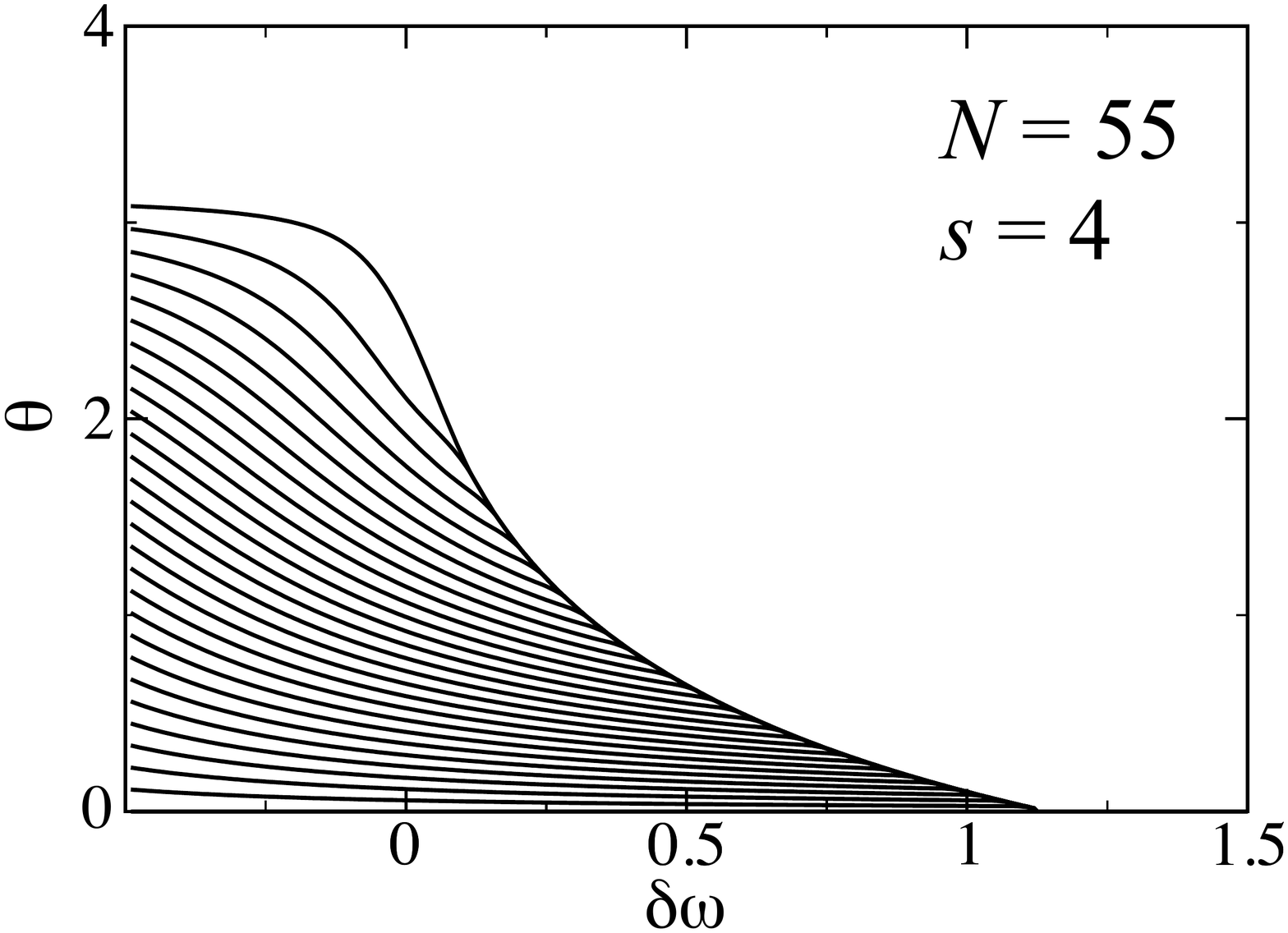}
\label{eigenvalue_finiteN_4}
}\end{minipage}
\caption{Numerical computations of the eigenvalues as a function of $\delta \omega = \omega -1/2$ in the model given in Eq.~(\ref{Z_GWW_V}) with $s=2,4$ and $N=55$. This demonstrates a pile up of eigenvalues when $s=4$, which motivated the ansatz of Eq.~(\ref{rho_even}).
}
\label{eigenval_fig}
\end{figure}

There is a qualitative difference between $s=4$ and all the other cases.
This difference was first discovered on the basis of numerical
analysis for $s=4$ and $N=55$ in Eq.~(\ref{Z_GWW_V}).  As can be seen in Fig.~(\ref{eigenval_fig}),
the eigenvalues are separate for $s=2$, while they pile up at the endpoints for $s=4$.  

This suggests that at infinite $N$,
the eigenvalues $\theta(x) $ with $\frac{x_0}{2} \leq x  \leq \frac{1}{2}$ become a single value $\theta_0=\theta(\frac{x_0}{2})$ for some $x_0$, and likewise for the other endpoint $-\theta_0$. Therefore, the expectation value of a function $f(\theta)$ can be written as
\begin{eqnarray}
\left< f(\theta) \right> 
&=& 
\int^{-\frac{x_0}{2}}_{-\frac{1}{2}} dx f(\theta(x)) + 
\int^{\frac{x_0}{2}}_{-\frac{x_0}{2}} dx f(\theta(x)) +
\int^{\frac{1}{2}}_{\frac{x_0}{2}} dx f(\theta(x)) 
\\
&=&
\frac{1-x_0}{2} \left( f(-\theta_0) + f(\theta_0) \right) + \int^{\theta_0}_{-\theta_0} d\theta \, \widetilde{\rho} (\theta) f(\theta) 
\label{f_theta_expectation}
\end{eqnarray}
where $\widetilde{\rho} (\theta)$ is a smooth function defined for the interval $-x_0/2 \leq x \leq x_0/2$. 
This ansatz corresponds to the following eigenvalue density
\begin{equation}
\rho(\theta) =
\frac{1-x_0}{2} \left\{ \delta(\theta-\theta_0)+\delta(\theta+\theta_0) \right\} + \widetilde{\rho} (\theta)  \, .
\label{rho_even}
\end{equation}
The two parameters $x_0$ and $\theta_0$ are related by the normalization condition, which can be derived by setting $f=1$ in Eq.~(\ref{f_theta_expectation}):
\begin{equation}
  x_0 = \int^{\theta_0}_{-\theta_0} d \theta \; \widetilde{\rho}(\theta) \; .
 \label{x0_even}
\end{equation}

In order to solve for $\widetilde{\rho}$, we take $s-1$ derivatives of the equation of motion (\ref{EoM_GWW_Integral_1}) with respect to $x$ for $-x_0/2 \leq x \leq x_0/2$:
\begin{equation}
\omega \cos \theta = \int^{\pi}_{-\pi} d\theta' \rho(\theta') \sum^{\infty}_{n=1} \cos \left(n \theta- n \theta' \right)  .
\end{equation}
Using the Poisson summation formula
\begin{equation}
\sum^\infty_{n=1} \cos( n \phi ) = - \frac{1}{2}+ \pi \; \delta(\phi) \, ,
\end{equation}
and the normalization condition
(\ref{normalization_condition}), we obtain
\begin{equation}
\widetilde{\rho}(\theta) = \frac{1}{2\pi} \left( 1 + 2 \omega \cos \theta \right) .
\label{rho_tilde}
\end{equation}
We now solve for the two unknowns, $\theta_0$ and $x_0$,
using the equation of motion (\ref{EoM_GWW_Integral_2}) with Eqs.~(\ref{rho_even}) and (\ref{rho_tilde}), and the normalization condition (\ref{x0_even}), which can be now written as
\begin{equation}
x_0 = \frac{\theta_0 +2 \, \omega \, \sin \theta_0}{\pi} 
\label{x0_p4}
\end{equation}
by using Eq.~(\ref{rho_tilde}).

To solve the equation of motion for $s=2$ we use the identity
\begin{equation}
\sum^{\infty}_{n=1} \frac{\sin (n \phi)}{n} =-\frac{1}{2}\phi+\frac{\pi}{2}
\mbox{sign} (\phi) 
\end{equation}
where $-2 \pi \leq \phi \leq 2 \pi$. 
This is proportional to the derivative of the Bernoulli polynomial $B_2(\left| \phi \right| / (2\pi))$ given in Eq.~(\ref{s_even_Fourier}). The equation of motion is satisfied for any value of $x_0$ in Eq.~(\ref{rho_even}). It turns out that the one minimizes the potential is when $x_0 =1$, $i.e.$, when there is no pile up of eigenvalues. This is consistent with the case of finite but large $N$ as shown in Fig.~(\ref{eigenvalue_finiteN_2}). Therefore, the eigenvalue density is 
\begin{equation}
\rho(\theta) = \widetilde{\rho} (\theta)= \frac{1}{2 \pi } \left(1+ 2 \omega \cos \theta \right)  
\label{rho_s2} 
\end{equation}
where $-\theta_0 \leq \theta \leq \theta_0$.
The is equivalent to the density below the GWW point (\ref{rho_belowGWW_omega}), except here it has a gap at the endpoints.  
This solution is consistent with the one found in \cite{Pisarski:2012bj}.
The normalization condition of Eq.~(\ref{x0_p4}) with $x_0=1$ gives
\begin{equation}
\omega = \frac{\pi-\theta_0}{2 \sin(\pi - \theta_0)}  \; ,
\label{boundary_p2}
\end{equation}
which implicitly defines the endpoint of the gap, $\theta_0(\omega)$.
From Eqs.~(\ref{rho_n_theta}) and (\ref{rho_s2}), the first Polyakov loop equals
\begin{equation}
  \rho_1 (\omega) = \frac{1}{\pi}
  \left\{ \sin\theta_0+ \omega \left(  \theta_0 + \cos \theta_0 \sin \theta_0 \right) \right\} \; . 
\label{rho1_s2}
\end{equation}
For small $\delta \omega = \omega - \frac{1}{2}>0$, 
\begin{equation}
  \rho_1(\omega) =
  \frac{1}{2} +\delta \omega - \frac{32 \sqrt{3}}{5 \pi} \delta \omega^{5/2} +\mathcal{O} (\delta \omega^{7/2}) \, .
\end{equation}

The eigenvalue density for $s=4$ is
\begin{equation}
\rho(\theta) =
  \frac{\pi - \theta_0 - 2 \omega \sin \theta_0}{2 \pi } \left\{ \delta(\theta-\theta_0)+\delta(\theta+\theta_0) \right\}
  +\frac{1}{2\pi} \left(1+2\omega \cos \theta \right) 
  \, 
\label{rho_s4}
\end{equation}
from Eqs.~(\ref{rho_even}), (\ref{rho_tilde}) and (\ref{x0_p4}). 
For $s=4$, the effective potential
involves the fourth Bernoulli polynomial.  The equation of motion
involves the third Bernoulli polynomial:
\begin{equation}
\sum^{\infty}_{n=1}\frac{ \sin (n \phi)}{n^3} = 
\frac{1}{12} \left(2\, \pi^2 \, \phi - 3\, \pi\, \mathrm{sign} (\phi) \, \phi^2 + \phi^3 \right) \; ,
\end{equation}
where $-2\pi \leq \phi \leq 2 \pi$.
The equation of motion (\ref{EoM_GWW_Integral_2}) with the above two equations gives 
\begin{equation}
0= \frac{\theta}{6 \pi} \left\{\left(\pi - \theta_0 \right)^3 - 6 \left(\pi-\theta_0 \right) \omega \cos \theta_0 - 6 \omega \sin \theta_0 \right\}
\; .
\end{equation}
It is satisfied for $-\theta_0 \leq \theta \leq \theta_0$ if 
\begin{equation}
\omega  = \frac{\left(\pi-\theta_0\right)^3 }{6 \sin \theta_0+6 \left(\pi-\theta_0\right) \cos \theta_0} \; .
\label{boundary_p4}
\end{equation}
This analytic form agrees with the largest eigenvalue for $N=55$ shown in Fig.~\ref{eigenvalue_finiteN_4}. 
When $s=4$, at $\omega=\pi^2/6$ the eigenvalue density collapses to a
$\delta$-function at a single point, $\theta=0$;
this also agrees with Fig.~\ref{eigenvalue_finiteN_4}. 
Using Eqs.~(\ref{rho_n_theta}) and (\ref{rho_s4}), we can write the first Polyakov loop for $s=4$ as
\begin{equation}
\rho_1 (\omega) = \frac{ \omega \, \theta_0+ \sin \theta_0 + \cos \theta_0 \left(\pi - \theta_0 - \omega \sin \theta_0 \right)}{\pi}
\label{rho1_s4}
\end{equation}
where $\theta_0(\omega)$ is given by Eq.~(\ref{boundary_p4}).
About the GWW point,
\begin{equation}
\rho_1(\omega) = \frac{1}{2} + \delta \omega - \, \frac{640 \, \sqrt{5}}{63 \, \pi} \; \delta \omega^{7/2} +
\mathcal{O}(\delta \omega^{9/2}) \, , 
\end{equation}
with $\delta \omega = \omega - \frac{1}{2}>0$.

\subsection{Near the GWW point}
\label{Summary_NearGWW}

\begin{figure}

\begin{minipage}{.5\textwidth}
\centering
\subfloat[$3$rd derivative of the free energy.]{\label{F_3rd}\includegraphics[scale=.46]{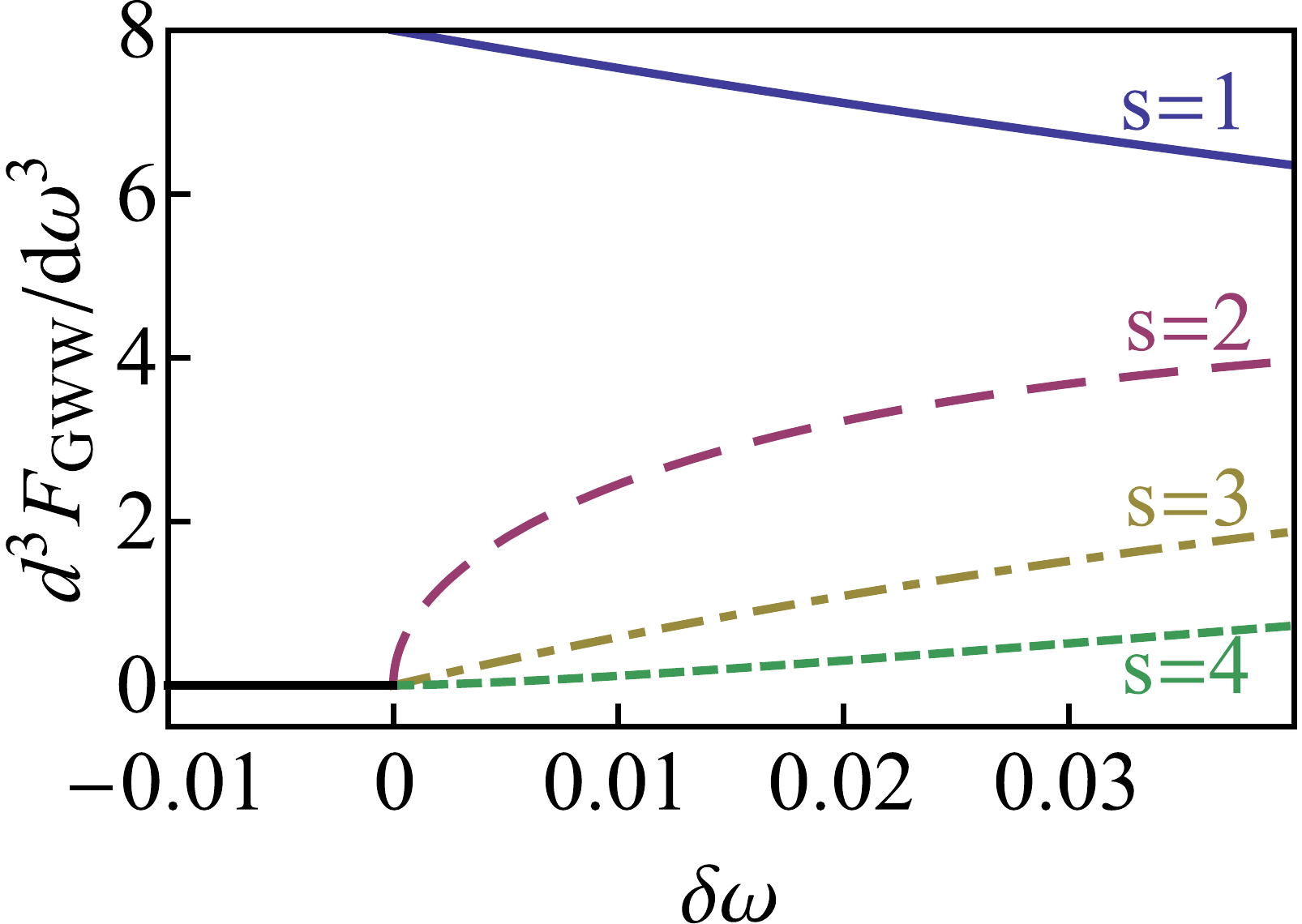}}
\end{minipage}%
\begin{minipage}{.5\textwidth}
\centering
\subfloat[$4$th derivative of the free energy.]{\label{F_4th}\includegraphics[scale=.4]{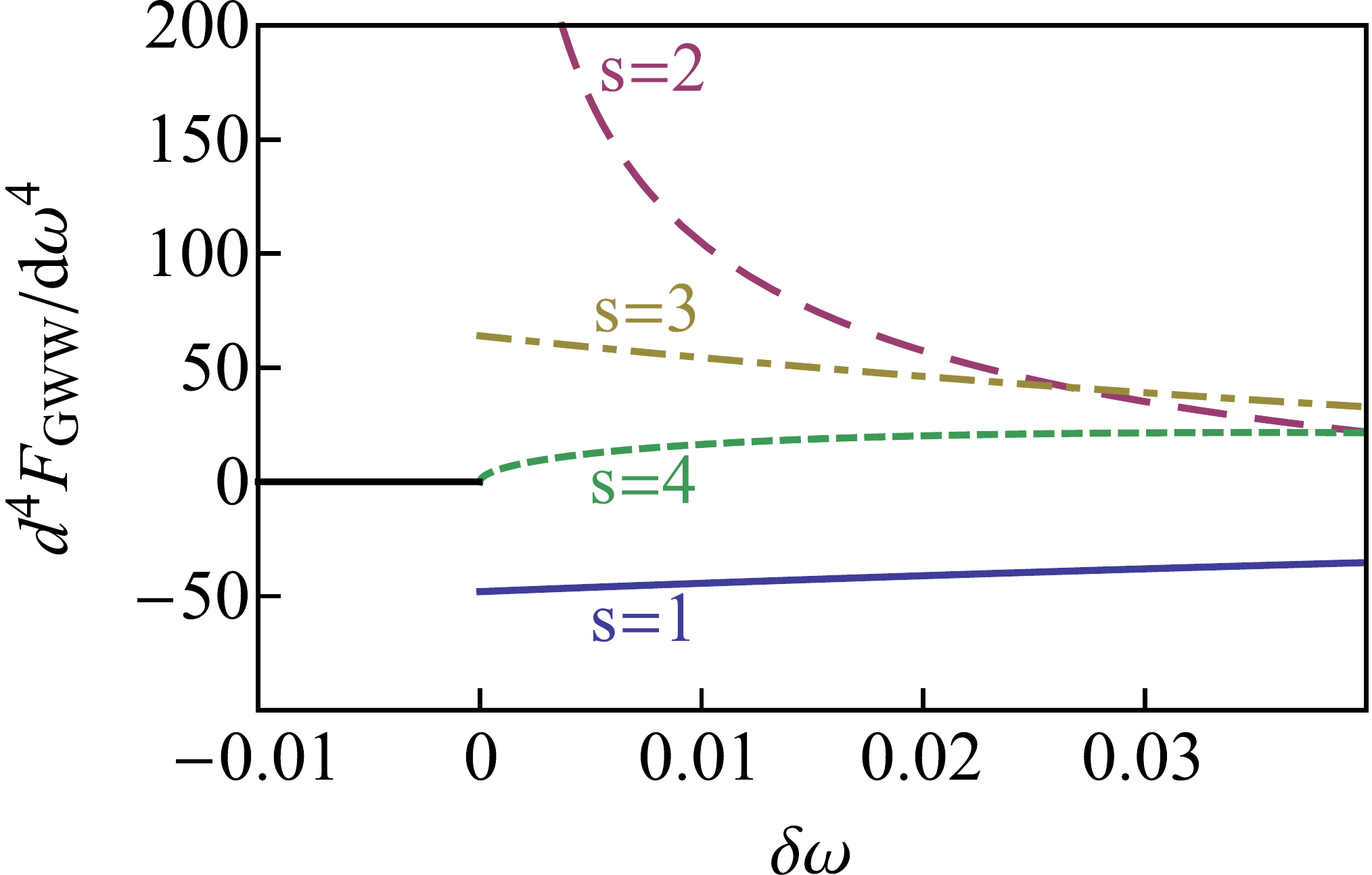}}
\end{minipage}\par\medskip
\centering
\subfloat[$5$th derivative of the free energy.]{\label{F_5th}\includegraphics[scale=.46]{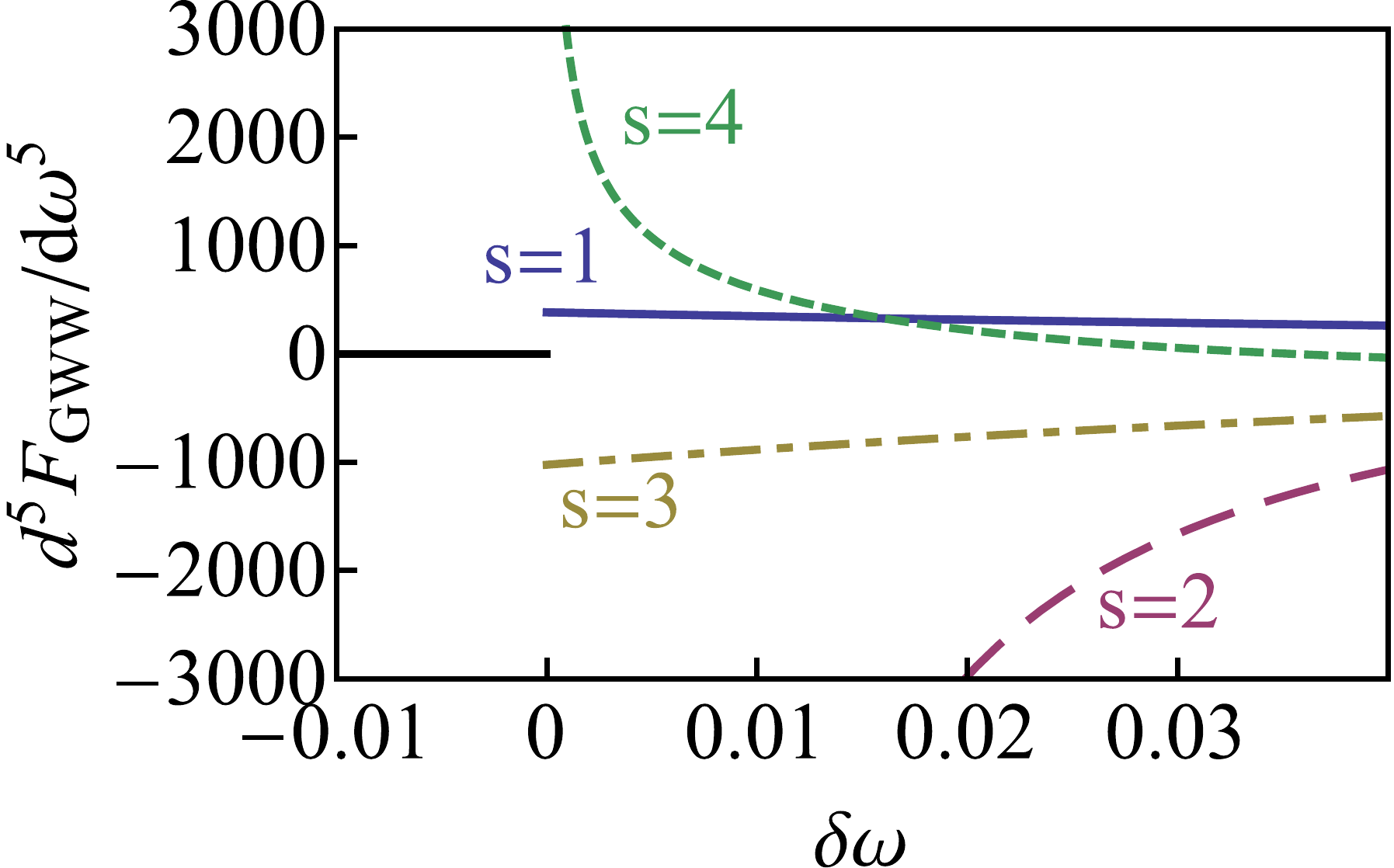}}
\caption{The derivatives of the free energy near the GWW point. The plot
  on the top left hand side shows that the transition for $s=1$ is of third
  order.  The plot on the top right hand side shows that $s=2$ and $3$ are
  transitions of fourth order.  The plot on the bottom shows that $s=4$ are transition of fifth order. }
\label{F_GWW_Derivatives}
\end{figure}

Given the solution for the effective potential
in Eq.~(\ref{V_GWW}) at infinite $N$ for $s=1,2,3$ and $4$, we can then use the eigenvalue
density to compute the free energy of Eq.~(\ref{Z_GWW_V}). 
This can be done analytically only for $s=1$, but in all cases
we can compute the free energy near the GWW point order by order in
$\delta \omega = \omega-\frac{1}{2}$ by using Eqs.~(\ref{rho1_s1}), (\ref{rho1_s2}),
(\ref{rho1_s3}), (\ref{rho1_s4}) for $s=1,2,3,4$, respectively, replacing
$h$ by $\omega$ in 
Eq.~(\ref{F_GWW_0}).  We find:
\begin{equation}
F_{\rm GWW}(\omega) = -\frac{1}{4} - \delta \omega - \delta \omega^2 + 
\left\{ 
\begin{array}{lcl}
0 & \mbox{for} & \delta \omega \leq 0
\\ 
v_s \delta \omega^{\left( 5+s \right)/2} + \mathcal{O}(\delta \omega^{\left(7+s \right)/2})  & \mbox{for} & \delta \omega > 0  \; ,
\end{array}
\right.
\label{F_GWW}
\end{equation}
where
\begin{equation}
  v_{1} = \frac{4}{3} \; , \;
  v_2 = \frac{128 \sqrt{3}}{35 \pi} \; , \;
  v_3 = \frac{8}{3} \; , \;
  v_4 = \frac{2560 \sqrt{5}}{567 \pi} \; .
  \label{vs}
\end{equation}
The order of the discontinuity with respect to $\omega$
depends upon $s$: as illustrated in Fig.~(\ref{F_GWW_Derivatives}), for $s=1$ the third
derivative is discontinuous; for $s=2$ and $3$, the fourth derivative; and
for $s=4$, the fifth derivative.  As discussed in the previous section, the
behavior of the free energy seen in Fig.~(\ref{F_GWW_Derivatives}) is unchanged by the quartic coupling as long as it is sufficiently small. 

\begin{figure}[t]\centering
\begin{minipage}{.47\textwidth}
\subfloat[$h=0$ and $b_1=0$.]{
\includegraphics[scale=0.42]{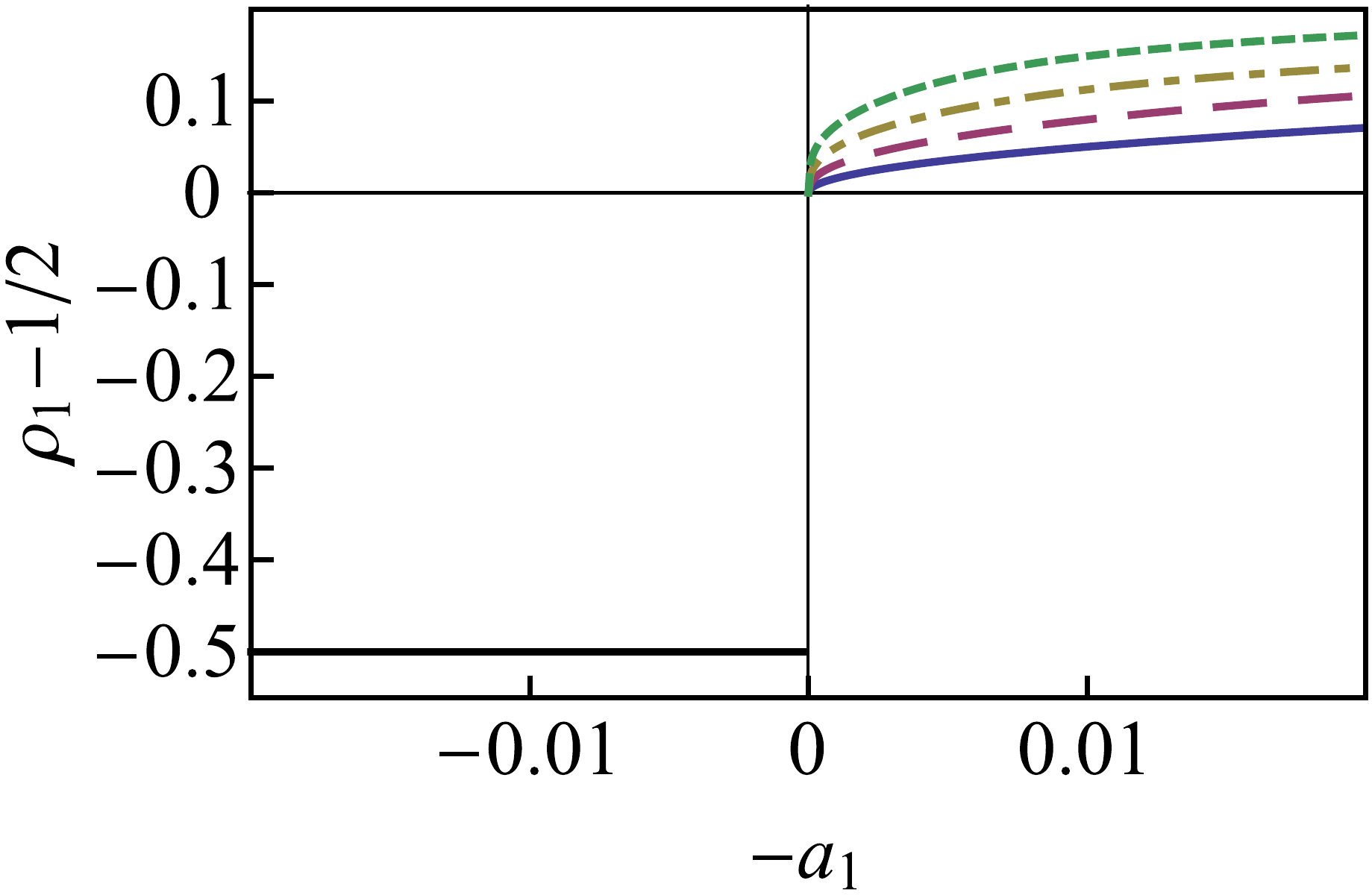}
\label{PolyakovLoop_1}
}\end{minipage}\; \;
\begin{minipage}{.47\textwidth}
\subfloat[$h=0.1$ and $b_1=0$.]{
\includegraphics[scale=0.42]{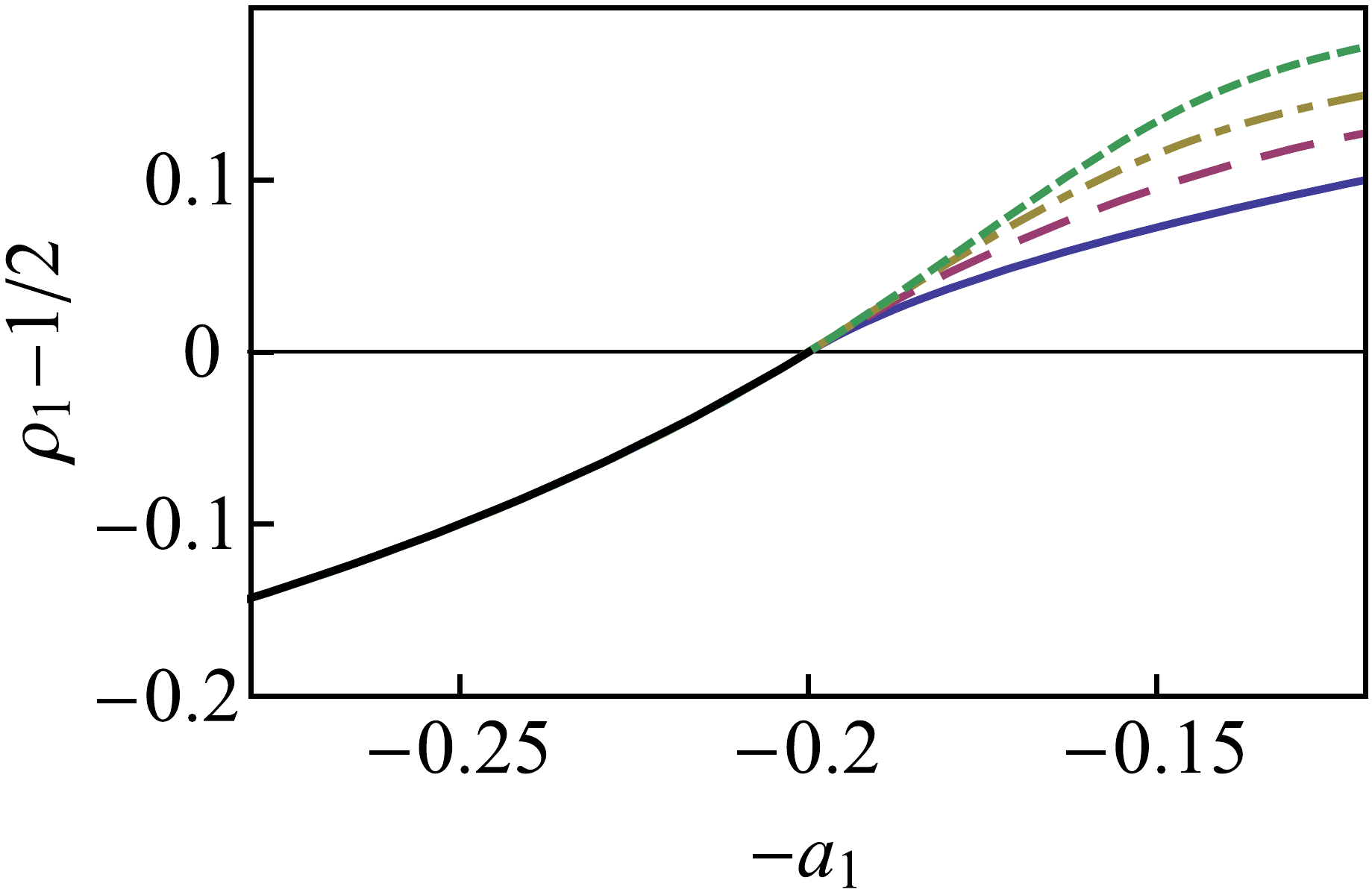}
\label{PolyakovLoop_2}
}\end{minipage}
\caption{The first Polyakov loop $\rho_1$ as a function of the coefficient of its quadratic term, $-a_1$.
  From bottom to top, $s=1,2,3,4$. }
\label{PolyakovLoop_zerob}
\end{figure}

\begin{figure}[t]\centering
\begin{minipage}{.47\textwidth}
\subfloat[$h=0$ and $b_1=-0.005$.]{
\includegraphics[width=\textwidth]{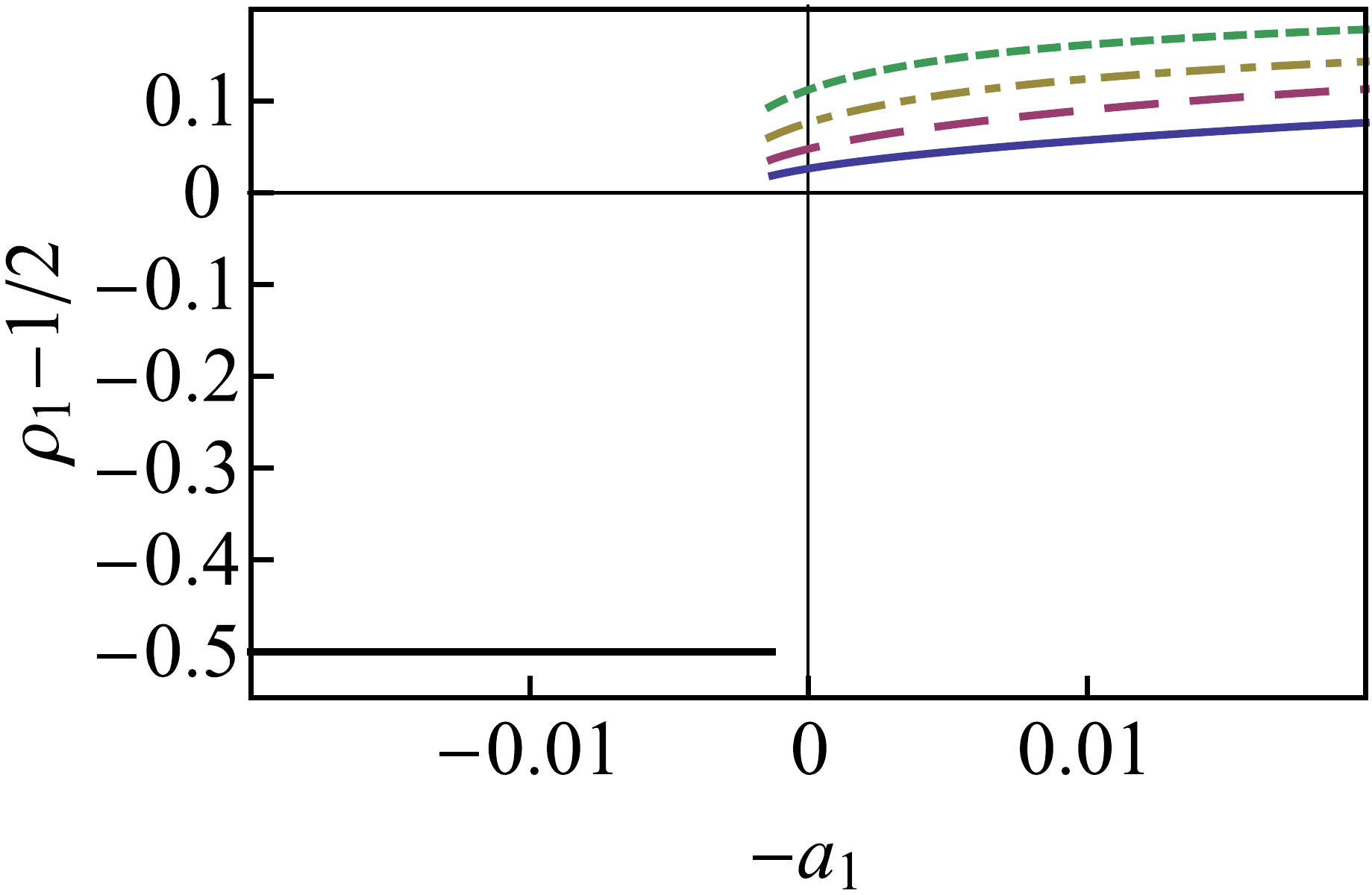}
\label{PolyakovLoop_3}
}\end{minipage}\; \;
\begin{minipage}{.47\textwidth}
\subfloat[At the first-order phase transition point.]{
\includegraphics[width=\textwidth]{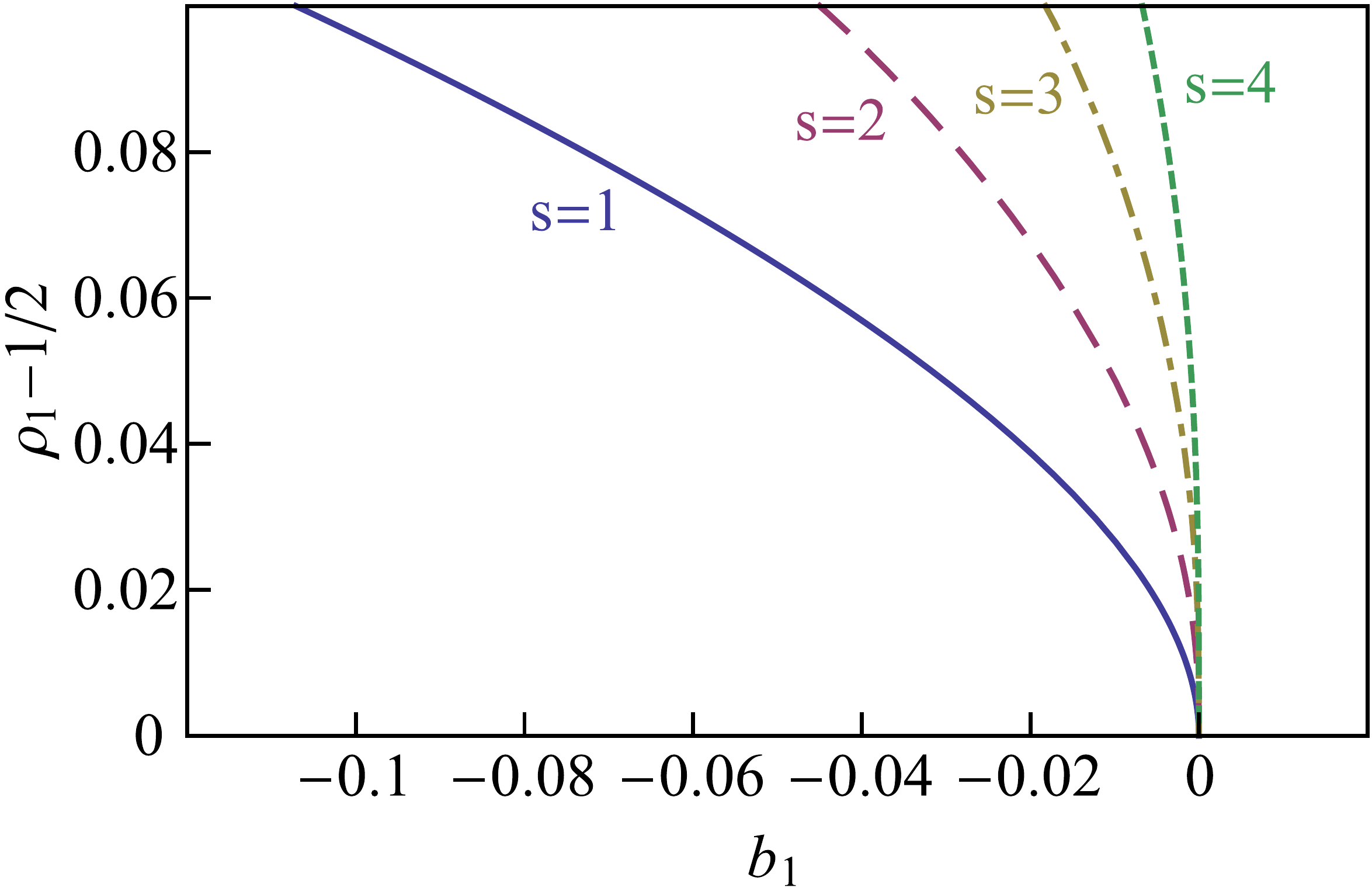}
\label{PolyakovLoop_b}
}\end{minipage}
\caption{The dependence of the first Polyakov loop $\rho_1$ on the quartic coupling $b_1$ in zero external field: from bottom to top, $s=1,2,3,4$.  }
\label{PolyakovLoop_negativeb}
\end{figure}

\begin{figure}[t]\centering
\includegraphics[scale=0.5]{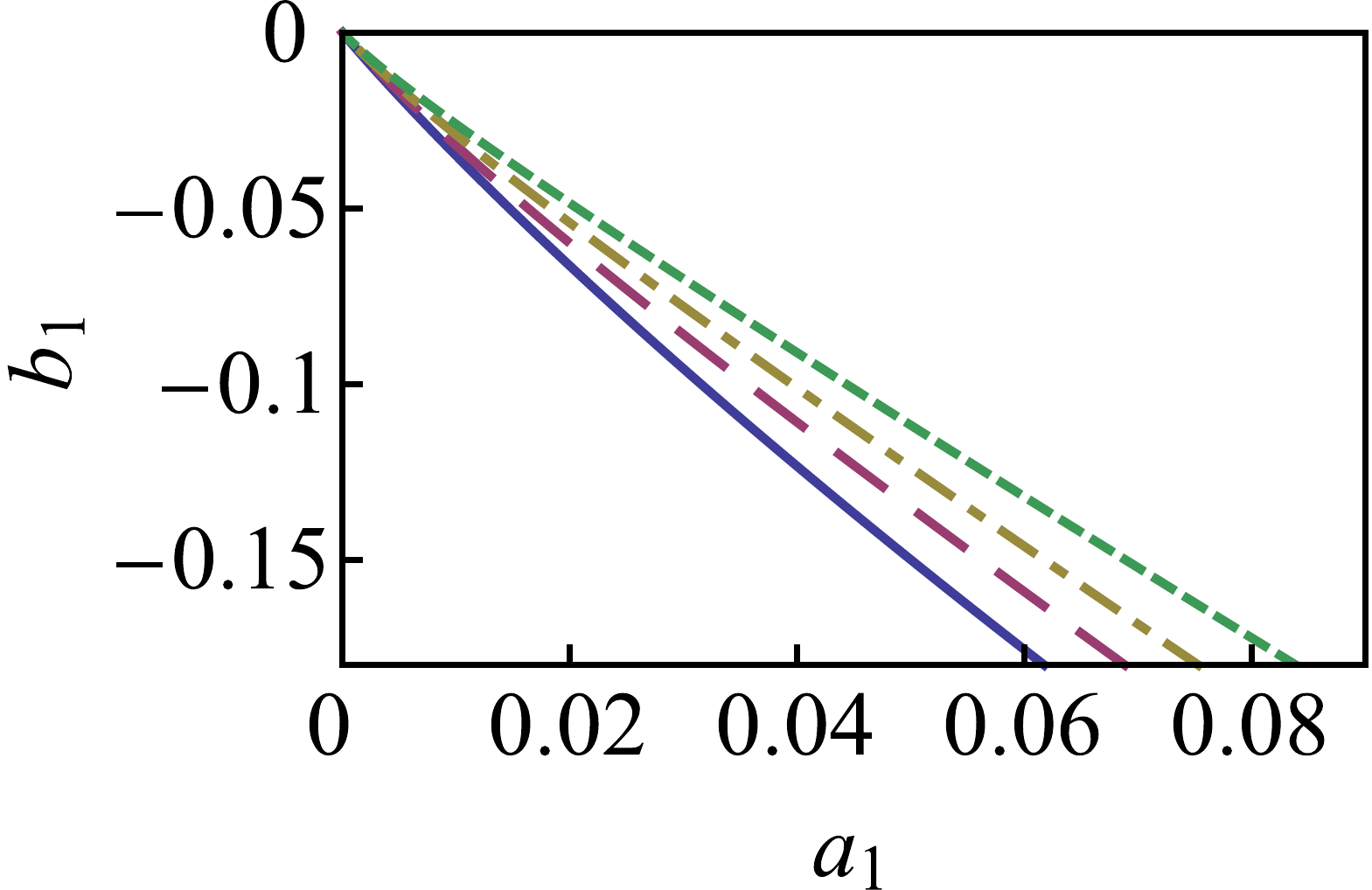}
\caption{The dependence of the first order line in zero field, $h=0$:
  from left to right, $s=1,2,3,$ and $4$. The blue solid line for $s=1$
	corresponds to the red dashed line in Fig.~(\ref{PhaseDiagram_2D}).
}
\label{1stCL}
\end{figure}

After inverting $\rho_1 (\omega)$ to obtain $\omega(\rho_1)$,
we can compute the Legendre transform of the potential by using
Eq.~(\ref{Gamma_GWW}). 
The behavior up to the leading nonanalytic term for $\delta \rho_1 = \rho_1 -1/2$ is
\begin{equation}
  \Gamma_{\rm GWW} (\rho_1) = \frac{1}{4} + \delta \rho_1 + \delta \rho^2_1
  + 
  \left\{ 
\begin{array}{lcl}
0 & \mbox{for} & \delta \rho_1 \leq 0
\\ 
v_s \delta \rho^{\left(5+s\right)/2}_1 + \mathcal{O}(\delta \rho^{\left(7+s\right)/2}_1)  & \mbox{for} & \delta \rho_1 > 0  \; ,
\end{array}
\right.
\label{Gamma_GWW_model}
\end{equation}
given the $v_s $ in Eq.~(\ref{vs}).  
Equation (\ref{Gamma}) allows one to calculate the Legendre
transform $\Gamma$ of the free energy as a function of $\rho_1$, $b_1$, $c_1$, $d_1$, and $h$.

In our model we set $d_1$ as a dimensionless constant and let $c_1$ change as a function
of $T/T_d$. Without loss of generality we can fix $d_1=1$,
\begin{equation}
-a_1 = 1-c_1(T/T_d) \, , 
\end{equation}
where $c_1$ is an unknown function of $T$.  Model studies of the deconfining
transition in $3+1$ \cite{Meisinger:2001cq, Dumitru:2010mj,*Dumitru:2012fw}
and $2+1$ dimensions \cite{Bicudo:2013yza, *Bicudo:2014cra} show that
the pressure, or equivalently the interaction measure, 
depends sensitively on $c_1$ .

To illustrate the physics we leave $c_1$ to be arbitrary and
assume that the mass term for the first Polyakov loop, $-a_1$,
is a monotonically increasing function of temperature.
As we discussed in
Sec.~\ref{Order_GWW}, when the quartic coupling $b_1 $ is zero or positive, the phase transition
occurs at $a_1=0$ and  $c_1 =1$, while negative coupling gives a phase
transition when $a_1>0$ and $c_1>1$.

We compute the first Polyakov loop, $\rho_1$, as a function of the
three parameters, $a_1$, $b_1$, and $h$.
Figure (\ref{PolyakovLoop_zerob}) shows the behavior of the first Polyakov loop
as a function of  $-a_1$, i.e. temperature, with zero quartic
coupling.  The figure on the left hand side is for zero external field, $h=0$: in
all cases, the first Polyakov loop jumps from $\rho_1 = 0$ to $\frac{1}{2}$ when
$a_1 = 0$.  This jump is typical of a first order phase transition.
As the temperature, or equivalently $- a_1$ increases, so does
the first Polyakov loop, with the increase greater for larger $s$.
This can be understood that the confining potential is weaker for larger
values of $s$.

In the presence of a nonzero external field all Polyakov loops
are nonzero, Fig.~(\ref{PolyakovLoop_2}). There is a GWW phase
transition at some value of $a_1 > 0$ when the value of the
first Polyakov loop $\rho_1 = \frac{1}{2}$.  At this point
there is {\it always} a transition of higher order, where the order depends
upon $s$, as discussed above.

Lastly we consider introducing a quartic coupling.  We assume it is
negative, as a positive coupling drives the transition to be of second order
about $\rho_1 = 0$.  This is certainly not supported by numerical simulations
on the lattice \cite{Lucini:2012gg, *Lucini:2013qja}.

In Fig.~(\ref{PolyakovLoop_negativeb}) we show the behavior for a small, negative
value of the quartic coupling $b_1$.  We assume that $|b_1|$ is small, so that
implicitly we are near the GWW point.  
Taking a fixed value of $b_1=-0.005$, $h=0$ and varying the quadratic
term in $\rho_1$, Fig.~(\ref{PolyakovLoop_3}) shows
that at the deconfining transition the first Polyakov loop jumps to a value
above $\frac{1}{2}$; the value that $\rho_1$ jumps to depends upon $s$.
The variation of $\rho_1 - \frac{1}{2}$ with $b_1$ is illustrated in 
Fig.~(\ref{PolyakovLoop_b}).

In Figure (\ref{1stCL}) we show how the position of the first order line changes
with $b_1$ in zero external field, $h=0$.  The blue solid line for $s=1$ corresponds to the red dashed line in Fig.~(\ref{PhaseDiagram_2D}). The model dependence of the first order transition line is small.

\section{Conclusions}
\label{Conclusions}

After discussing the most general effective potential for Polyakov loops in
Sec.~\ref{Effective_potential}, in Sec.~\ref{Phase_structure} we showed
that {\it if} double trace terms dominate the potential, then one is
naturally led to a phase diagram in which the generalized Gross-Witten-Wadia (GWW)
transition, whose order is larger than second, is ubiquitous.  From Sec.~\ref{Effective_potential}, there
is no generic reason why double trace terms should dominate.  However,
as we discussed there, there are several cases in which, rather unexpectedly, they do.

We then solved the models of  Eq.~(\ref{V_eff_model}) for $s = 1 ,2, 3$ and $4$
in Sec.~\ref{Models}.  We considered only simple forms of the coefficients for the double trace terms, $a_n = c_n - d_n$, where the positive (negative) contribution is responsible for the (de)confined phase. These models are illustrative, and not
representative.  For example, in $3+1$ dimensions, a term with $d_n \sim 1/n^4$
arises perturbatively \cite{Gross:1980br}.  It is also necessary to add a second,
``confining'' term, such as $c_n \sim 1/n^2$
\cite{Dumitru:2010mj, *Dumitru:2012fw, Kashiwa:2012wa, Pisarski:2012bj, *Lin:2013qu,Meisinger:2001cq}.
In order to generate a deconfining transition,
by necessity the sign of the confining term must be opposite to that of the
perturbative term.
In $2+1$ dimensions, a term with $d_n \sim 1/n^3$ is similarly generated
perturbatively \cite{KorthalsAltes:1996xp}.
From numerical simulations on the lattice \cite{Caselle:2011fy, *Caselle:2011mn},
matrix models with $c_n \sim 1/n^2$ are also natural \cite{Bicudo:2013yza, *Bicudo:2014cra}.

Our model in Sec.~\ref{Models} contains both the confining terms $c_n$ and the perturbative terms $-d_n$, but the latter consists only of the first term $-d_1$ for the first Polyakov loop.  
The matrix model with the full perturbative
terms $d_n \sim 1/n^d$ with $d=4$ and the confining potential $c_n \sim 1/n^s$ with $s=2$, relevant to $3+1$ dimensions, was solved at large $N$ \cite{Pisarski:2012bj, *Lin:2013qu}.
By comparing to the free energy for $s=2$ in Eq.~(\ref{F_GWW})
with that of Ref.~\cite{Pisarski:2012bj, *Lin:2013qu}, one finds the it is identical up to the leading non-analytic term, $v_2 \, \delta \omega^{7/2}$.  This is expected because the full coefficients $a_n = c_n -d_n$ can be approximated as $a_n \sim c_n $ for $n \geq 2$ below $T_d$.  Therefore we expect that for the matrix model with both the confining and full perturbative terms, there is a region in the confined phase in Fig.~\ref{PhaseDiagram_2D} where the approximation $c_n - d_n \sim c_n$ for $s \leq d$ and $1<n$ is valid, and thus our exact solution is a good approximation for the the full potential. It would
be interesting to check if this is indeed the case for the models relevant to $2+1$ dimensions.

Our results show that the nature of phase transition depends sensitively upon how
close the theory is to a model with only double trace terms.   We studied
this by adding a quartic term for the first Polyakov loop, Eq.~(\ref{V_eff_model}).
Lattice simulations for pure Yang-Mills theory at large $N$ indicates
that the deconfining phase transition is of first order \cite{Lucini:2012gg, *Lucini:2013qja}.
This implies that the quartic coupling is either zero or negative.
As shown in Sec.~\ref{Models}, at the GWW point the expectation value of
the first Polyakov loop equals $\frac{1}{2}$ at $T_d^+$.  Numerical
simulations on the lattice find a result
close to this value \cite{Mykkanen:2012ri}, which suggests
that the theory at large $N$ is close to the GWW point.
This could be tested by adding an external field for the first
Polyakov loop and measuring the free energy and its derivative
as the external field is varied.
As seen in Figs.~(\ref{Distribution}) and (\ref{F_GWW_Derivatives}), these
quantities change dramatically about the GWW point.
Alternately, one could look for phase transitions as the
lattice coupling is varied ~\cite{Bursa:2005tk}.  

\begin{figure}[t]\centering
\begin{minipage}{.48\textwidth}
\subfloat[$N = N_f  = 3$]{
\includegraphics[height=0.62\textwidth]{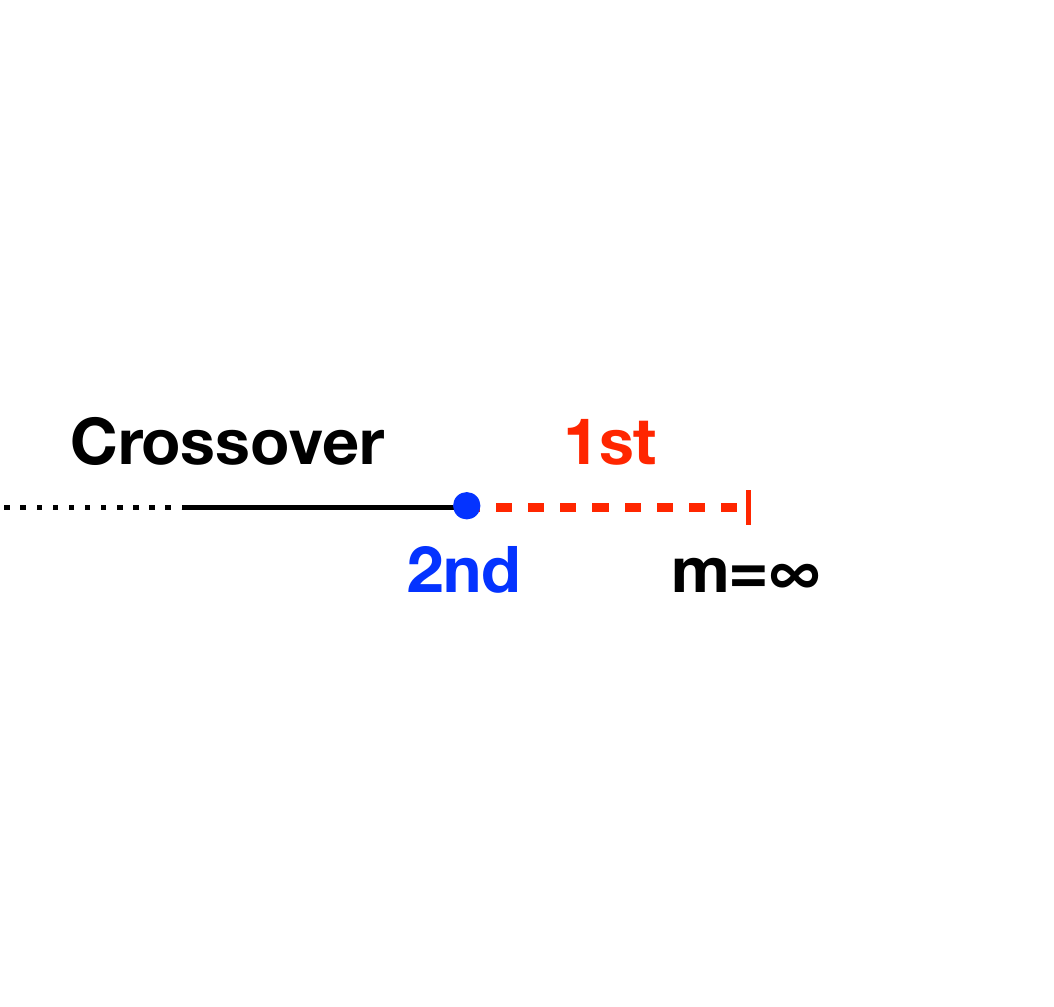}
\label{Columbia_Finite_N}
}\end{minipage}\; \;
\begin{minipage}{.48\textwidth}
\subfloat[$N = N_f  = \infty$]{
\includegraphics[height=0.62\textwidth]{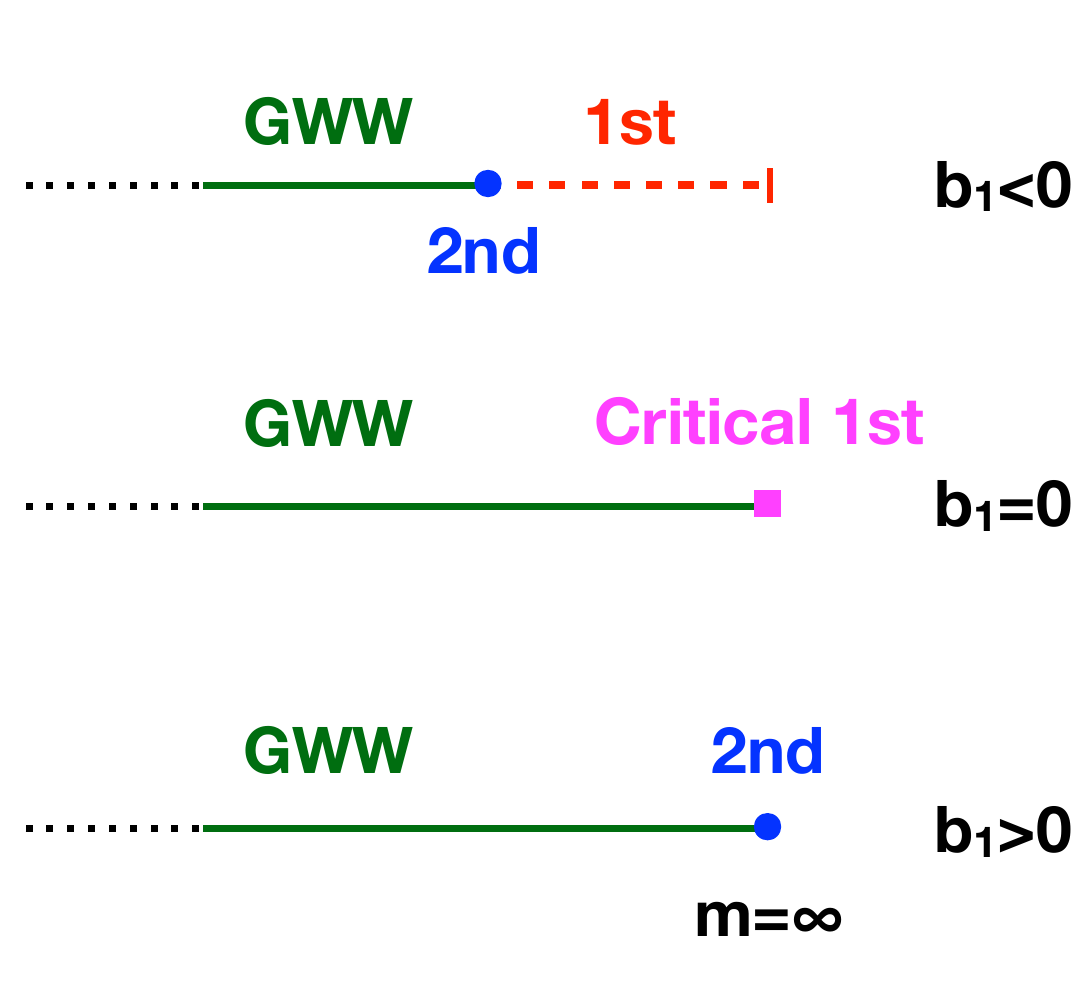}
\label{Columbia_Infinite_N}
}\end{minipage}
\caption{Phase diagram as a function of heavy quark mass $m$. For an infinite
numbers of colors and flavors, the phase diagram depends on the value of quartic coupling $b_1$ for the first Polyakov loop $\rho_1$.}
\label{PhaseDiagram_HeavyQuark}
\end{figure}

Since heavy quarks act like a background magnetic field for the first Polyakov loop \cite{Schnitzer:2004qt},
adding $N_f$ flavors of heavy quarks, with $N_f \sim N \rightarrow \infty$,
also changes the phase diagram in characteristic
ways.  For three colors and three flavors the Columbia phase diagram \cite{Lucini:2012gg, *Lucini:2013qja}
implies that as the quark mass increases, a crossover becomes a first order transition.
As illustrated in Fig.~(\ref{PhaseDiagram_HeavyQuark}), for intermediate quark masses,
where there is a crossover for $N_f = N = 3$, there must be a {\it line} of GWW
transitions.

If the quartic coupling for the first Polyakov loop, $b_1$, is positive, one ends with a second
order transition for infinitely heavy quarks.  If $b_1$ is negative, there is a line of
first order transitions for sufficiently heavy quarks.

What is especially interesting is the third possibility: $b_1$, and {\it all} associated couplings
from three or more traces, vanish.  In that case, the line of GWW transitions continues to
infinite quark masses and ends with the critical first order.  That is, that the {\it only} terms which contribute to the effective
potential are those with double trace terms, $i_1 = - i_2$ and $j_1 = j_2 = 1$
in Eq.~(\ref{general_zn_potential}).

Such a limitation on the possible terms does not follow merely from the global symmetry of $Z(N)$,
but must be a larger symmetry special to infinite $N$.  {\it If} this happens, and the deconfining
transition is critical first order at infinite $N$, then even though the transition is of first order, one has
a conformally symmetric theory at $T_d^+$.  For an ordinary second order transition,
continuity implies that the critical exponents, {\it etc.}, are the same on either side of
the phase transition.   In the present case, as the energy density
and order parameter are discontinuous at $T_d$,
it is even possible that there is a {\it different} conformally symmetric theory
at $T_d^-$.  This cannot be
studied in our models, since the free energy is of order $\sim N^2$ above $T_d$,
and only $\sim 1$ below.

While base speculation, gauge theories are objects of singular beauty, especially in the
limit of infinite $N$.

\acknowledgments

We thank M. Anber, A. Cherman, A.~Dumitru,  K.~Fukushima, F.~Karsch, S.~Sharma, Y.~Tanizaki  for useful
discussions.  H.~N. and V.~S. are supported by the Special Postdoctoral
Researchers program of RIKEN.  R.D.P. thanks the U.S. Department of Energy for support
under contract DE-SC0012704.

\appendix

\section{Large $N$ perturbation theory in the presence of the background field}
\label{AppendixA}

It is interesting to consider whether the simple structure of Eq.~(\ref{pert_two_loop})
persists to higher loop order.  It is well known that because of infrared divergences
in $3+1$ dimensions at nonzero temperature, that the free energy is a power series
not in $g^2$, but in $\sqrt{g^2}$.  At nonzero temperature, gluons have
euclidean energies $= 2 \pi T n$, where $n$ is an integer, $n = 0, \pm 1, \pm 2 \ldots$.
Static modes with $n = 0$ have zero energy at tree level.  At one loop
order the static modes develop a thermal (Debye) mass $\sim g T$.
Integration over these modes in three spatial dimensions gives a term
in the free energy $\sim T (g T)^3 \sim g^3 T^4$.  Beyond $\sim g^3$,
higher order corrections to the free energy are $\sim g^4, g^5$, {\it etc.}~\footnote{Additionally, 
there is a logarithmic dependence on $g$.}

This power counting changes in the presence of a background field for the thermal
Wilson line.
For a constant background field $A_0^{i j} \sim T \, \theta^i \delta^{i j}/g$, gluon modes in the adjoint
representation have euclidean energies $\sim T (2 \pi n + \theta_i - \theta_j) $.
Consequently, assuming a general background field with 
$\theta_i \neq \theta_j$, the energy of off-diagonal gluons is always nonzero, even if $n = 0$.
In contrast, diagonal gluons are insensitive to the background field and
have modes with zero energy.

Consider, however, the free energy in the limit of large $N$.  There are $\sim N^2$ off-diagonal
gluons, and only $\sim N$ diagonal gluons.  Thus only off-diagonal gluons contribute to the
term in the free energy $\sim N^2$, and for this term
the free energy is a power series in $g^2 N$.  This assumes, of course,
that the $\theta_i$ are not small, $|\theta_i| > g$.

It would be useful to compute the effective potential for a thermal
Wilson line to three loop order at large $N$, $\sim g^4 N^2$ \cite{nishimura_skokov}.
The leading terms at large $N$ can, but need not, include terms with
four traces.
As we saw in this paper, terms with four traces, $a_{(1,-1)}^{(2,2)} =b_1 \neq 0$, greatly affect the properties of the deconfining phase transition.

\section{Alternative form of the free energy $F_{\rm GWW}$}
\label{AppendixB}

In this appendix, we show another way to compute the free energy for the GWW
potential based on the paper \cite{Brezin:1977sv}.  The potential given in
Eq.~(\ref{V_GWW_Sum}) can be written in terms of the eigenvalue density as
\begin{equation}
  V_{\rm GWW} = \int d\theta \; \rho(\theta)
  \int d\theta'  \; \rho(\theta')  \sum^{\infty}_{n=1}
  a_n \cos(n (\theta-  \theta')) - 2 h \int d\theta \; \rho(\theta) \; \cos \theta \; 
\label{V_GWW_Appendix}
\end{equation}
where $a_n >0$. 
The equation of motion can be found by taking a functional derivative $\delta
V_{\rm GWW} /\delta \theta(x)$ as in Sec.~\ref{Solution_GWW}:
\begin{eqnarray}
0
=
\int d\theta' \; \rho(\theta')  \sum^{\infty}_{n=1}  n \, a_n \, \sin(n (\theta - \theta')) - h \, \sin \theta \; .
\end{eqnarray}
We integrate it with respect to $\theta$,
\begin{equation}
C= \int d\theta' \; \rho(\theta')  \; \sum^{\infty}_{n=1}  a_n \, \cos(n (\theta -  \theta')) - h \, \cos \theta  \; ,
\label{EoM_Appendix}
\end{equation}
where $C$ is a constant.  By setting  $\theta =0$, we have
\begin{equation}
C = \int d\theta' \; \rho(\theta')  \; \sum^{\infty}_{n=1}  a_n \, \cos(n \theta') - h \; .
\end{equation}
Substituting this into Eq.~(\ref{EoM_Appendix}) and 
integrating it with $\int d \theta \rho(\theta)$, we have
\begin{equation}
  \int d\theta\;  \rho(\theta)   \int d\theta' \; \rho(\theta')  \;
  \sum^{\infty}_{n=1}  a_n  \cos(n (\theta -  \theta'))  
=
\int d\theta \rho(\theta)  \sum^{\infty}_{n=1}  a_n  \cos(n \, \theta) + h \left( \rho_1 - 1\right) \; .
\end{equation}
Here $\rho$ is the solution for the equation of motion. 
Therefore using this expression into Eq.~(\ref{V_GWW_Appendix}), we obtain the free energy
\begin{equation}
F_{\rm GWW}(h) = \int d\theta \; \rho(\theta)  \; \sum^{\infty}_{n=1}  a_n  \cos(n\theta)- h \left( \rho_1 + 1\right) 
= \sum^\infty_{n=1} \, a_n \, \rho_n - \, h \left( \rho_1 +1 \right) \; .
\label{F_GWW_1}
\end{equation}
This agrees with Eqs.~(\ref{F_GWW_0}) and (\ref{F_GWW_2}) when $V_{\rm eff}$ is given as in Eq.~(\ref{V_GWW_Sum}).

\bibliography{GWW} 
\newpage

U.S. Department of Energy Office of Nuclear Physics or High Energy Physics

{\it Notice:}
This manuscript has been co-authored by employees of Brookhaven
Science Associates, LLC under Contract No. DE-SC0012704 with
the U.S. Department of Energy. The publisher by accepting the manuscript for
publication acknowledges that the United States Government retains a
non-exclusive, paid-up, irrevocable, world-wide license to publish or
reproduce the published form of this manuscript, or allow others to do so,
for United States Government purposes.
This preprint is intended for publication in a journal or proceedings.
Since changes may be made before publication, it may not be cited or
reproduced without the author’s permission.

{\it DISCLAIMER}:
This report was prepared as an account of work sponsored by an agency of the
United States Government.  Neither the United States Government nor any
agency thereof, nor any of their employees, nor any of their contractors,
subcontractors, or their employees, makes any warranty, express or implied,
or assumes any legal liability or responsibility for the accuracy,
completeness, or any third party’s use or the results of such use of any
information, apparatus, product, or process disclosed, or represents that
its use would not infringe privately owned rights. Reference herein to any
specific commercial product, process, or service by trade name, trademark,
manufacturer, or otherwise, does not necessarily constitute or imply its
endorsement, recommendation, or favoring by the United States Government or
any agency thereof or its contractors or subcontractors.  The views and
opinions of authors expressed herein do not necessarily state or reflect
those of the United States Government or any agency thereof.

\end{document}